\pdfoutput=1

\documentclass[]{raa}           

\usepackage{graphicx,times}
\usepackage{amssymb,amsmath}
\usepackage{color}

\begin{document}

\title{Spectroscopic determination of C, N, and O abundances 
of solar-analog stars based on the lines of hydride molecules}

\volnopage{ {\bf 20XX} Vol.\ {\bf X} No. {\bf XX}, 000--000}

\setcounter{page}{1}

\author{Yoichi Takeda}
\institute{11-2 Enomachi, Naka-ku, Hiroshima-shi, 730-0851, Japan; {\it ytakeda@js2.so-net.ne.jp}}

\vs \no
{\small Received 2022 November 10; accepted 2022 December 11}

\abstract{
Photospheric C, N, and O abundances of 118 solar-analog stars were determined
by applying the synthetic-fitting analysis to their spectra in the blue or 
near-UV region comprising lines of CH, NH, and OH molecules, with an aim 
of clarifying the behaviors of these abundances in comparison with [Fe/H]. 
It turned out that, in the range of $-0.6 \la$~[Fe/H]~$\la +0.3$, 
[C/Fe] shows a marginally increasing tendency with decreasing [Fe/H] 
with a slight upturn around [Fe/H]~$\sim 0$, [N/Fe] tends to somewhat
decrease towards lower [Fe/H], and [O/Fe] systematically increases (and thus 
[C/O] decreases) with a decrease in [Fe/H]. While these results are qualitatively 
consistent with previous determinations mostly based on atomic lines, the 
distribution centers of these [C/Fe], [N/Fe], and [O/Fe] at the near-solar metallicity
are slightly negative by several hundredths dex, which is interpreted as due to 
unusual solar abundances possibly related to the planetary formation of our solar system.
However, clear anomalies are not observed in the [C,N,O/Fe] ratios of planet-host stars.
Three out of four very Be-deficient stars were found to show anomalous 
[C/Fe] or [N/Fe] which may be due to mass transfer from the evolved companion,
though its relation to Be depletion mechanism is still unclear.
}

\authorrunning{Y. Takeda}            
\titlerunning{CNO abundances of solar-analog stars}  
\maketitle

\twocolumn

%
 
\section{Introduction} 

Carbon (C), nitrogen (N), and oxygen (O) are representatively 
abundant light elements in the Universe next to hydrogen and helium. 
They are synthesized or burned inside of a star and expelled at the last stage 
of stellar evolution out to the galactic gas, but the way how this process
undergoes in which kind of stars is different for each element.
While O is produced and distributed mainly by short-lived high-mass stars,
longer-lived low-to-intermediate mass stars may also contribute to the 
enrichment of C or N. Therefore, CNO abundances in low-mass main-sequence 
stars like our Sun can be a clue to studying the chemical evolution 
of these elements, because such stars have diversified ages and the 
composition of galactic gas at the time of star formation are retained
in their atmospheres. Specifically, the runs of [C/Fe], [N/Fe], and [O/Fe]
with a change of [Fe/H]\footnote{
As usual, [X/H] is the differential abundance for element X of a star 
relative to the Sun defined as [X/H] $\equiv$ $A_{*}$(X) $-$ $A_{\odot}$(X), 
where $A$(X) is the logarithmic number abundance of element X (normalized 
with respect to H as $A$(H) = 12). Likewise, the notation [X/Y] is defined 
as [X/Y] $\equiv$ [X/H] $-$ [Y/H].}
(representative of metallicity) in solar-type stars\footnote{
Although no definite classification scheme exists, the following terminology  
may hold as a rule of thumb: 
(1) ``Solar-type stars'' are late-type stars of solar associates in the broad sense
(e.g., early K through late F-type dwarfs or subgiants).
(2) ``Solar analogs'' are early G-type dwarfs which have properties analogous 
to the Sun (e.g., differences in $T_{\rm eff}$ and $\log g$ are  
$\pm \la$~100--200~K and $\pm \la$~0.1--0.2~dex).
(3) ``Solar twins'' apply to a special group of stars with parameters 
very resembling the Sun (e.g., $\Delta T_{\rm eff}$ and $\Delta \log g$ 
are within a few tens K and within a few hundredths dex). 
} play an important role in this context. 

Given this astrophysical significance, not a few spectroscopic determinations 
of CNO abundances for solar-type stars have been carried out so far (e.g., 
Takeda \& Honda 2005, and the references therein). However, it does not seem 
necessarily easy to accomplish a sufficient precision. For example, in the 
work of Takeda \& Honda (2005), C or O abundances derived from permitted 
and forbidden lines were not in satisfactory agreement. Actually, several 
disadvantages are involved in using lines of neutral atoms (C~{\sc i}, N~{\sc i}, 
O~{\sc i}) usually adopted: (i) Usable lines are rather few in number and 
do not have sufficient strengths. (ii) Many are high-excitation lines and 
thus considerably dependent upon $T$ (temperature) in late-type stars. 
(iii) Though forbidden lines ([C~{\sc i}] or [O~{\sc i}]) are inert to $T$, 
they are so weak and apt to be contaminated by blending. (iv) Stronger lines 
existing in the near-IR region tend to suffer a considerable non-LTE effect.

In the meantime, another possibility of CNO abundance determinations is to 
make use of lines of hydride molecules (CH, NH, and OH) in the blue or near-UV 
region, which however has not been mainstream, because this task comes with 
several difficulties: (i) These lines are in rather unfavorable wavelength 
regions (accessibility, line crowdness; etc.). (ii) Since the classical
analysis using line-by-line equivalent widths is hardly practicable, it is 
requisite to compare the observed and theoretically synthesized spectra.
(iii) Because of low dissociation potentials, the populations of 
these molecules are quite sensitive to $T$, which means that resulting 
abundances are appreciably dependent upon adopted atmospheric models.
(iv) More seriously, abundances ($A$) derived from these molecular lines
(especially those of UV region such as NH) tend to suffer more or less 
systematic errors for unknown reasons, as shown by the pioneering work 
of Laird (1985). 

However, although such systematic errors involved in $A$(X) (X = C, N, O) 
derived from hydride molecules may be unavoidable, differential abundances 
relative to the Sun ([X/H]; cf. footnote~1) may still be acceptable at least 
for solar-type stars, because errors tend be cancelled if stars are not much 
different from the Sun. Actually, several studies revealed that reasonable 
results of [X/H] could be obtained for FGK-type dwarfs (including planet-host
stars) by applying the spectrum-synthesis analysis to molecular line features 
in the blue--UV region; e.g., Ecuvillon et al. (2006) for OH, 
Su\'{a}rez-Andr\'{e}s et al. (2016) for NH, and Su\'{a}rez-Andr\'{e}s et al. 
(2017) for CH. 

If so, it must be more preferable to carry out a differential analysis relative 
to the Sun exclusively for ``solar-analog'' stars (Sun-like early G-type dwarfs;
cf. footnote~2), by which systematic errors would be largely suppressed.
Therefore, sufficient reliability is expected for the differential abundances 
([X/H]) of solar analogs; moreover, high precision would be accomplished
by averaging the results derived from a number of spectral regions (thanks to
the availability of numerous lines for these molecules).  

The author's group previously carried out comprehensive investigations for 
118 solar-analog stars in comparison with the Sun, which were published in a 
series of papers: Takeda et al. (2007) (stellar parameters and Li abundances), 
Takeda et al. (2010) (stellar activity estimated from Ca~{\sc ii} 8542 and 
its relation to rotation), Takeda et al. (2011) (behaviors of Be abundances
determined from Be~{\sc ii} 3131), and Takeda et al. (2012) (detection of 
low-level activity using Ca~{\sc ii} 3934).
In connection with this project, we acquired high-dispersion spectra of near-UV 
through blue region ($\sim$~3000--4600~\AA) for all of the sample stars, which 
were employed for the latter two studies. Since these spectra are just suitable 
for doing the differential analysis of molecular lines mentioned above, I decided 
to conduct new CNO abundance determinations of these solar analogs based on 
CH, NH, and OH lines, while paying attention to the following points.
\begin{itemize}
\item
How are the behaviors of [C/Fe], [N/Fe], and [O/Fe] with a change of [Fe/H] 
(e.g., gradient, dispersion, zero-point) within $\pm$ several tenths dex 
around [Fe/H] $\sim 0$?  It is interesting to compare their trends with 
those previously derived from atomic lines.
The variation of [C/O] with [Fe/H] is also an important checkpoint.    
\item
Do stars harboring giant planets show any appreciable difference in terms of 
CNO abundances in comparison with the sample of non-planet-host stars?
\item
Takeda et al. (2011) serendipitously found 4 extraordinary Sun-like stars, 
in which Be is drastically deficient by $\ga 2$~dex compared to the others 
(and the Sun). It is interesting to check whether these Be-depleted stars 
exhibit any peculiarities in CNO abundances, which may provide some information
on the origin of Be anomaly. 
\end{itemize}
The purpose of this article is to describe the outcome of this analysis.

\section{Program stars}

\subsection{Atmospheric parameters}

The sample of 118 solar-analog stars in the solar neighborhood are the same 
as those adopted in Takeda et al. (2007), which were selected by the criterion 
of $0.62 \la B-V \la 0.67$ and $4.5 \la M_{V} \la 5.1$ (centered
around the solar values of $(B-V)_{\odot} = 0.65$ and 
$M_{V\odot} = 4.82$). Likewise, regarding the atmospheric 
parameters [$T_{\rm eff}$ (effective temperature), $\log g$ (surface gravity), 
$v_{\rm t}$ (microturbulence), and [Fe/H] (Fe abundance relative to the solar 
Fe abundance of $A_{\odot}$(Fe) = 7.50)] of these stars along with the Sun, 
those spectroscopically determined by Takeda et al. (2007) (standard solutions; 
cf. sect.~3.1.1 therein) were used unchanged.\footnote{Exceptionally, 
the parameters for HIP~41484 derived in Takeda et al. (2007) were wrong, 
because irrelevant observational data were erroneously used for this star. 
The correct parameters were later redetermined 
in Takeda et al. (2010) (cf. appendix~A therein).} 
The list of 118 stars (+ Sun) and their atmospheric parameters are presented 
in Table~1, where 12 planet-host stars\footnote{According to 
``The Extrasolar Planets Encyclopaedia'' site (http://exoplanet.eu/).
This number has significantly increased since the time of Takeda et al. (2011),
where only 5 stars out of these 118 solar analogs were regarded as
planet-harboring stars (cf. sect.~4.1.3 therein).}
and 4 Be-depleted stars (cf. sect.~4.2 in Takeda et al. 2011) are also indicated.
The atmospheric models are the same as adopted in our previous papers 
(cf. sect.~4.1 of Takeda et al. 2007), which were generated by 3-dimensionally 
interpolating Kurucz's (1993a) ATLAS9 grid of model atmospheres in terms of 
$T_{\rm eff}$, $\log g$, and [Fe/H] (metallicity). 

\subsection{Errors in $T_{\rm eff}$, $\log g$, and $v_{\rm t}$}

As described in sect.~3.1.1 of Takeda et al. (2007), these atmospheric parameters 
were determined by using the TGVIT program (Takeda et al. 2002, 2005) based 
on the equivalent widths of Fe~{\sc i} and Fe~{\sc ii} lines, while requiring 
three conditions to be simultaneously satisfied:  
(a) independence of $A$(Fe) upon $\chi_{\rm low}$, 
(b) independence of $A$(Fe) upon the equivalent widths, 
and (c) matching the mean abundances of $\langle A$(Fe~{\sc i})$\rangle$ 
and $\langle A$(Fe~{\sc ii})$\rangle$.
Since solar $gf$ values (cf. sect.~2 in Takeda et al. 2005) were adopted in this 
application to solar-analog stars, the statistical errors involved in 
$T_{\rm eff}$, $\log g$, and $v_{\rm t}$ (estimated by the procedure described 
in sect.~5.2 of Takeda et al. 2002) are sufficiently small, which are typically 
on the order of $\sim \pm 20$~K, $\sim \pm 0.05$~dex, and $\sim \pm 0.1$~km~s$^{-1}$, 
respectively. See also sect.~3.4 in Takeda et al. (2011); the errors derived for 
individual stars are given in electronic table~E3 of that paper.

\subsection{Observational data}

The high-dispersion spectra of 118 program stars and Vesta (substitute 
for the Sun) covering $\sim$~3000--4600~\AA\ with a resolving power of 
$R \simeq 60000$ used in this study are those obtained in 2009--2010 with the 
High Dispersion Spectrograph (HDS) placed at the Nasmyth platform of the 8.2-m 
Subaru Telescope atop Mauna Kea (see sect.~2 of Takeda et al. 2011 for more details).


\setcounter{table}{0}
\begin{table*}[h]
\begin{center}
\caption{Stellar parameters and the CNO abundance results of 118 program stars.}
\setlength{\tabcolsep}{5.0pt}
\small
\begin{tabular}{ccccccccccc}
\hline\hline
Star & HIP & $T_{\rm eff}$ & $\log g$ & $v_{\rm t}$ & [Fe/H] & [C/H] & [N/H] & [O/H] & [C/O] & Remark \\
No.  & number & (K) & (dex) & (km~s$^{-1}$) & (dex) & (dex) & (dex) & (dex) & (dex) &   \\ 
\hline\noalign{\smallskip}
  1 & 001499 & 5724 & 4.45 & 0.95 & +0.20 & +0.20 & +0.23 & +0.09 & +0.12 & PHS\\
  2 & 001598 & 5693 & 4.33 & 0.96 & $-$0.27 & $-$0.38 & $-$0.38 & $-$0.24 & $-$0.14 & \\
  3 & 001803 & 5817 & 4.41 & 1.17 & +0.24 & +0.11 & +0.20 & +0.10 & +0.01 & \\
  4 & 004290 & 5719 & 4.40 & 1.10 & $-$0.12 & $-$0.21 & $-$0.20 & $-$0.16 & $-$0.04 & \\
  5 & 005176 & 5855 & 4.39 & 1.03 & +0.19 & +0.18 & +0.22 & +0.14 & +0.04 & \\
  6 & 006405 & 5728 & 4.38 & 0.96 & $-$0.14 & $-$0.24 & $-$0.28 & $-$0.14 & $-$0.09 & \\
  7 & 006455 & 5716 & 4.57 & 0.99 & $-$0.09 & $-$0.15 & $-$0.18 & $-$0.09 & $-$0.06 & \\
  8 & 007244 & 5755 & 4.52 & 1.12 & $-$0.04 & $-$0.12 & $-$0.12 & $-$0.09 & $-$0.03 & \\
  9 & 007585 & 5784 & 4.50 & 1.04 & +0.07 & +0.00 & $-$0.02 & +0.01 & +0.00 & \\
 10 & 007902 & 5613 & 4.39 & 0.91 & $-$0.01 & $-$0.05 & $-$0.17 & $-$0.02 & $-$0.03 & \\
 11 & 007918 & 5841 & 4.30 & 1.12 & +0.01 & $-$0.03 & +0.00 & $-$0.01 & $-$0.02 & \\
 12 & 008486 & 5805 & 4.45 & 1.13 & $-$0.06 & $-$0.22 & $-$0.23 & $-$0.13 & $-$0.09 & \\
 13 & 009172 & 5763 & 4.55 & 1.12 & +0.06 & $-$0.08 & $-$0.05 & $-$0.02 & $-$0.06 & \\
 14 & 009349 & 5788 & 4.35 & 1.07 & +0.01 & $-$0.03 & +0.01 & +0.00 & $-$0.03 & \\
 15 & 009519 & 5853 & 4.45 & 1.22 & +0.14 & $-$0.04 & $-$0.04 & +0.02 & $-$0.06 & PHS \\
 16 & 009829 & 5579 & 4.25 & 0.94 & $-$0.31 & $-$0.36 & $-$0.44 & $-$0.31 & $-$0.06 & \\
 17 & 010321 & 5707 & 4.60 & 1.04 & +0.00 & $-$0.10 & $-$0.11 & $-$0.09 & $-$0.01 & \\
 18 & 011728 & 5708 & 4.40 & 1.02 & +0.02 & $-$0.05 & $-$0.03 & $-$0.05 & +0.00 & \\
 19 & 012067 & 5709 & 4.41 & 0.96 & +0.20 & +0.16 & +0.18 & +0.09 & +0.06 & \\
 20 & 014614 & 5726 & 4.26 & 1.00 & $-$0.12 & $-$0.18 & $-$0.19 & $-$0.11 & $-$0.07 & \\
 21 & 014623 & 5742 & 4.52 & 1.09 & +0.12 & +0.03 & +0.08 & +0.01 & +0.02 & \\
 22 & 015062 & 5735 & 4.49 & 0.94 & $-$0.29 & $-$0.38 & $-$0.40 & $-$0.23 & $-$0.15 & \\
 23 & 015442 & 5682 & 4.50 & 0.87 & $-$0.19 & $-$0.25 & $-$0.29 & $-$0.16 & $-$0.10 & \\
 24 & 016405 & 5738 & 4.32 & 1.03 & +0.26 & +0.27 & +0.25 & +0.17 & +0.10 & \\
 25 & 017336 & 5671 & 4.55 & 0.94 & $-$0.13 & $-$0.21 & $-$0.01 & $-$0.11 & $-$0.10 & BED \\
 26 & 018261 & 5873 & 4.43 & 0.97 & +0.02 & $-$0.06 & $-$0.03 & +0.04 & $-$0.10 & \\
 27 & 019793 & 5828 & 4.51 & 1.26 & +0.19 & +0.08 & +0.15 & +0.05 & +0.03 & \\
 28 & 019911 & 5672 & 4.34 & 1.10 & $-$0.13 & $-$0.33 & $-$0.37 & $-$0.32 & $-$0.02 & \\
 29 & 019925 & 5767 & 4.53 & 0.99 & +0.07 & +0.01 & +0.03 & +0.02 & $-$0.01 & \\
 30 & 020441 & 5771 & 4.42 & 1.10 & +0.13 & +0.10 & +0.17 & +0.05 & +0.05 & \\
 31 & 020719 & 5831 & 4.36 & 1.24 & +0.13 & +0.02 & +0.13 & $-$0.02 & +0.04 & \\
 32 & 020741 & 5797 & 4.37 & 1.20 & +0.16 & +0.09 & +0.17 & +0.05 & +0.04 & \\
 33 & 020752 & 5923 & 4.46 & 1.13 & +0.16 & +0.03 & +0.05 & +0.06 & $-$0.03 & \\
 34 & 021165 & 5760 & 4.28 & 0.99 & $-$0.16 & $-$0.26 & $-$0.31 & $-$0.15 & $-$0.11 & \\
 35 & 021172 & 5625 & 4.27 & 0.90 & $-$0.10 & $-$0.16 & $-$0.27 & $-$0.08 & $-$0.07 & \\
 36 & 022203 & 5740 & 4.33 & 1.07 & +0.13 & +0.05 & +0.12 & +0.00 & +0.05 & \\
 37 & 023530 & 5601 & 4.36 & 0.91 & $-$0.24 & $-$0.18 & $-$0.18 & $-$0.11 & $-$0.07 & \\
 38 & 025002 & 5729 & 4.47 & 1.07 & $-$0.08 & $-$0.23 & $-$0.26 & $-$0.21 & $-$0.02 & \\
 39 & 025414 & 5635 & 4.49 & 0.89 & +0.10 & +0.09 & +0.06 & +0.01 & +0.08 & \\
 40 & 025670 & 5759 & 4.55 & 0.88 & +0.10 & +0.04 & +0.06 & +0.08 & $-$0.04 & \\
 41 & 026381 & 5518 & 4.47 & 0.87 & $-$0.45 & $-$0.47 & $-$0.65 & $-$0.31 & $-$0.16 & PHS \\
 42 & 027435 & 5697 & 4.45 & 0.93 & $-$0.22 & $-$0.27 & $-$0.28 & $-$0.18 & $-$0.09 & PHS \\
 43 & 029432 & 5712 & 4.32 & 1.00 & $-$0.12 & $-$0.14 & $-$0.16 & $-$0.11 & $-$0.03 & PHS \\
 44 & 031965 & 5770 & 4.31 & 0.99 & +0.05 & +0.02 & +0.00 & +0.03 & $-$0.01 & \\
 45 & 032673 & 5724 & 4.57 & 0.95 & +0.06 & +0.06 & +0.23 & +0.01 & +0.05 & BED \\
 46 & 033932 & 5891 & 4.38 & 1.10 & $-$0.12 & $-$0.20 & $-$0.22 & $-$0.07 & $-$0.13 & \\
 47 & 035185 & 5793 & 4.19 & 1.35 & +0.00 & $-$0.14 & $-$0.08 & $-$0.07 & $-$0.08 & \\
 48 & 035265 & 5804 & 4.37 & 1.04 & $-$0.02 & $-$0.05 & $-$0.04 & $-$0.04 & $-$0.02 & \\
 49 & 036512 & 5718 & 4.49 & 0.89 & $-$0.09 & $-$0.13 & $-$0.17 & $-$0.08 & $-$0.04 & \\
 50 & 038647 & 5714 & 4.43 & 0.95 & +0.01 & $-$0.14 & $-$0.15 & $-$0.06 & $-$0.07 & \\
 51 & 038747 & 5804 & 4.42 & 1.05 & +0.06 & $-$0.08 & $-$0.08 & +0.00 & $-$0.08 & \\
 52 & 038853 & 5899 & 4.27 & 1.03 & $-$0.05 & $-$0.12 & $-$0.14 & $-$0.05 & $-$0.07 & \\
 53 & 039506 & 5600 & 4.24 & 0.83 & $-$0.62 & $-$0.69 & $-$0.92 & $-$0.40 & $-$0.29 & \\
 54 & 039822 & 5758 & 4.35 & 0.90 & $-$0.22 & $-$0.26 & $-$0.32 & $-$0.12 & $-$0.14 & \\
 55 & 040118 & 5541 & 4.45 & 0.84 & $-$0.42 & $-$0.43 & $-$0.63 & $-$0.29 & $-$0.14 & \\
 56 & 040133 & 5698 & 4.33 & 0.97 & +0.12 & +0.05 & +0.05 & +0.03 & +0.02 & \\
 57 & 041184 & 5705 & 4.43 & 1.51 & +0.11 & $-$0.09 & +0.06 & $-$0.10 & +0.02 & \\
 58 & 041484 & 5864 & 4.33 & 0.92 & +0.05 & +0.02 & +0.02 & +0.07 & $-$0.06 & \\
 59 & 041526 & 5801 & 4.27 & 0.98 & $-$0.02 & $-$0.10 & $-$0.14 & $-$0.02 & $-$0.08 & \\
 60 & 042333 & 5816 & 4.44 & 1.08 & +0.14 & +0.05 & +0.10 & +0.06 & $-$0.01 & \\
 61 & 042575 & 5675 & 4.40 & 0.96 & +0.06 & +0.00 & +0.00 & +0.00 & +0.00 & \\
 62 & 043297 & 5691 & 4.46 & 1.05 & +0.08 & +0.01 & +0.02 & +0.02 & $-$0.01 & \\
 63 & 043557 & 5805 & 4.42 & 1.05 & $-$0.06 & $-$0.06 & $-$0.07 & $-$0.05 & $-$0.01 & \\
 64 & 043726 & 5769 & 4.47 & 1.01 & +0.11 & +0.09 & +0.16 & +0.03 & +0.06 & \\
 65 & 044324 & 5888 & 4.45 & 1.09 & $-$0.01 & $-$0.08 & $-$0.11 & $-$0.02 & $-$0.06 & \\
\hline
\end{tabular}
\end{center}
\end{table*}

\setcounter{table}{0}
\begin{table*}[h]
\begin{center}
\caption{(Continued.)}
\setlength{\tabcolsep}{5.0pt}
\small
\begin{tabular}{ccccccccccc}
\hline\hline
Star & HIP & $T_{\rm eff}$ & $\log g$ & $v_{\rm t}$ & [Fe/H] & [C/H] & [N/H] & [O/H] & [C/O] & Remark \\
No.  & number & (K) & (dex) & (km~s$^{-1}$) & (dex) & (dex) & (dex) & (dex) & (dex) &   \\ 
\hline\noalign{\smallskip}
 66 & 044997 & 5696 & 4.54 & 0.75 & +0.04 & $-$0.03 & +0.01 & +0.02 & $-$0.05 & \\
 67 & 045325 & 5935 & 4.47 & 0.97 & +0.18 & +0.17 & +0.22 & +0.22 & $-$0.06 & \\
 68 & 046903 & 5746 & 4.40 & 1.11 & $-$0.03 & $-$0.10 & $-$0.10 & $-$0.06 & $-$0.04 & \\
 69 & 049580 & 5782 & 4.41 & 0.87 & +0.02 & $-$0.05 & $-$0.05 & +0.03 & $-$0.08 & \\
 70 & 049586 & 5786 & 4.42 & 1.06 & +0.20 & +0.15 & +0.18 & +0.12 & +0.03 & \\
 71 & 049728 & 5744 & 4.40 & 0.98 & $-$0.07 & $-$0.09 & $-$0.12 & $-$0.06 & $-$0.03 & \\
 72 & 049756 & 5720 & 4.28 & 0.99 & +0.02 & $-$0.02 & $-$0.01 & $-$0.02 & +0.00 & \\
 73 & 050505 & 5590 & 4.44 & 0.84 & $-$0.17 & $-$0.23 & $-$0.29 & $-$0.19 & $-$0.04 & \\
 74 & 051178 & 5801 & 4.47 & 0.87 & $-$0.17 & $-$0.20 & $-$0.23 & $-$0.12 & $-$0.08 & \\
 75 & 053721 & 5819 & 4.19 & 1.15 & $-$0.02 & $-$0.05 & $-$0.01 & $-$0.03 & $-$0.02 & PHS \\
 76 & 054375 & 5803 & 4.37 & 0.96 & +0.14 & +0.03 & +0.06 & +0.05 & $-$0.02 & \\
 77 & 055459 & 5812 & 4.36 & 1.03 & +0.07 & +0.03 & +0.02 & +0.05 & $-$0.02 & \\
 78 & 055868 & 5757 & 4.49 & 0.95 & $-$0.15 & $-$0.25 & $-$0.27 & $-$0.12 & $-$0.12 & \\
 79 & 059589 & 5654 & 4.51 & 0.70 & $-$0.01 & $-$0.04 & $-$0.16 & +0.04 & $-$0.09 & \\
 80 & 059610 & 5829 & 4.34 & 1.04 & $-$0.06 & $-$0.09 & $-$0.09 & $-$0.05 & $-$0.04 & PHS \\
 81 & 062175 & 5683 & 4.19 & 0.90 & +0.13 & +0.02 & +0.00 & +0.01 & +0.01 & \\
 82 & 062816 & 5804 & 4.43 & 0.97 & +0.06 & $-$0.04 & +0.00 & +0.00 & $-$0.04 & \\
 83 & 063048 & 5655 & 4.32 & 0.91 & $-$0.02 & $-$0.01 & $-$0.11 & +0.00 & $-$0.02 & \\
 84 & 063636 & 5799 & 4.52 & 1.10 & $-$0.01 & $-$0.05 & $-$0.08 & $-$0.04 & +0.00 & \\
 85 & 064150 & 5726 & 4.42 & 0.99 & +0.05 & +0.05 & +0.03 & +0.00 & +0.05 & BED \\
 86 & 064747 & 5710 & 4.42 & 0.93 & $-$0.18 & $-$0.20 & $-$0.26 & $-$0.10 & $-$0.09 & \\
 87 & 070319 & 5678 & 4.42 & 0.96 & $-$0.33 & $-$0.35 & $-$0.50 & $-$0.24 & $-$0.11 & PHS \\
 88 & 072604 & 5655 & 4.24 & 0.84 & $-$0.14 & $-$0.19 & $-$0.30 & $-$0.07 & $-$0.11 & \\
 89 & 075676 & 5772 & 4.44 & 0.88 & $-$0.08 & +0.09 & $-$0.05 & $-$0.07 & +0.17 & BED \\
 90 & 076114 & 5709 & 4.42 & 1.02 & $-$0.02 & $-$0.05 & $-$0.05 & $-$0.05 & +0.00 & \\
 91 & 077749 & 5836 & 4.61 & 1.14 & +0.22 & +0.08 & +0.14 & +0.09 & $-$0.02 & \\
 92 & 078217 & 5749 & 4.43 & 1.10 & $-$0.22 & $-$0.32 & $-$0.33 & $-$0.17 & $-$0.16 & \\
 93 & 079672 & 5768 & 4.40 & 0.96 & +0.04 & $-$0.03 & $-$0.02 & +0.00 & $-$0.03 & \\
 94 & 085042 & 5676 & 4.48 & 0.99 & +0.03 & $-$0.03 & $-$0.06 & $-$0.03 & +0.01 & \\
 95 & 085810 & 5856 & 4.46 & 1.08 & +0.15 & +0.11 & +0.16 & +0.10 & +0.01 & \\
 96 & 088194 & 5693 & 4.33 & 0.98 & $-$0.08 & $-$0.15 & $-$0.18 & $-$0.09 & $-$0.06 & PHS \\
 97 & 088945 & 5800 & 4.38 & 1.44 & $-$0.01 & $-$0.19 & $-$0.05 & $-$0.17 & $-$0.02 & \\
 98 & 089282 & 5833 & 4.22 & 1.00 & +0.00 & $-$0.11 & $-$0.14 & $-$0.05 & $-$0.06 & \\
 99 & 089474 & 5755 & 4.20 & 1.04 & +0.01 & $-$0.01 & $-$0.01 & +0.01 & $-$0.02 & PHS \\
100 & 089912 & 5846 & 4.38 & 1.24 & +0.04 & $-$0.11 & $-$0.08 & $-$0.06 & $-$0.05 & \\
101 & 090004 & 5607 & 4.42 & 0.85 & $-$0.01 & $-$0.04 & $-$0.17 & +0.03 & $-$0.07 & PHS \\
102 & 091287 & 5648 & 4.46 & 0.88 & $-$0.01 & $-$0.06 & $-$0.11 & $-$0.06 & +0.01 & \\
103 & 096184 & 5863 & 4.45 & 1.00 & +0.13 & +0.12 & +0.13 & +0.12 & +0.00 & \\
104 & 096395 & 5816 & 4.48 & 1.00 & $-$0.10 & $-$0.15 & $-$0.21 & $-$0.09 & $-$0.06 & \\
105 & 096402 & 5661 & 4.20 & 1.00 & $-$0.03 & +0.02 & $-$0.08 & +0.03 & +0.00 & \\
106 & 096901 & 5742 & 4.32 & 1.01 & +0.08 & +0.07 & +0.04 & +0.05 & +0.02 & PHS \\
107 & 096948 & 5725 & 4.36 & 1.07 & +0.07 & +0.04 & +0.04 & $-$0.01 & +0.04 & \\
108 & 097420 & 5780 & 4.42 & 1.04 & +0.05 & $-$0.07 & $-$0.06 & +0.01 & $-$0.08 & \\
109 & 098921 & 5810 & 4.50 & 1.19 & +0.17 & +0.07 & +0.10 & +0.05 & +0.02 & \\
110 & 100963 & 5779 & 4.46 & 0.98 & +0.00 & $-$0.04 & $-$0.05 & $-$0.03 & $-$0.01 & \\
111 & 104075 & 5881 & 4.37 & 1.08 & +0.05 & $-$0.09 & $-$0.07 & $-$0.03 & $-$0.06 & \\
112 & 109110 & 5835 & 4.51 & 1.11 & +0.07 & $-$0.05 & $-$0.03 & +0.00 & $-$0.05 & \\
113 & 110205 & 5708 & 4.28 & 1.08 & $-$0.23 & $-$0.24 & $-$0.25 & $-$0.18 & $-$0.06 & \\
114 & 112504 & 5741 & 4.34 & 1.00 & +0.01 & $-$0.09 & $-$0.10 & $-$0.07 & $-$0.02 & \\
115 & 113579 & 5759 & 4.21 & 1.44 & +0.05 & $-$0.09 & +0.09 & $-$0.05 & $-$0.04 & \\
116 & 113989 & 5506 & 4.38 & 0.74 & $-$0.46 & $-$0.54 & $-$0.73 & $-$0.38 & $-$0.16 & \\
117 & 115715 & 5684 & 4.14 & 1.05 & $-$0.19 & $-$0.30 & $-$0.34 & $-$0.22 & $-$0.08 & \\
118 & 116613 & 5869 & 4.49 & 1.11 & +0.16 & +0.05 & +0.12 & +0.09 & $-$0.04 & \\
--- &   Sun  & 5761 & 4.43 & 1.00 & $-$0.01 & --- &   --- & ---   & ---     & \\
\hline
\end{tabular}
\end{center}
Following the sequentially assigned star No. (Column~1) and 
Hipparcos catalogue number (Column~2), four atmospheric parameters
(effective temperature, surface gravity, microturbulence, Fe abundance relative
to the solar abundance of $A_{\odot}$(Fe) = 7.50) taken from Takeda et al. (2007; 
standard solutions) are given in Columns~3--6. The CNO abundance results derived 
in this study ([C/H], [N/H], [O/H], [C/O]) are presented in Column~7--10. 
In Column~11, 12 planet-host stars and 4 Be-depleted stars are remarked as 
``PHS'' and ``BED'', respectively.
The parameters for the Sun are given in the last row.
\end{table*}

\section{Abundance determination}

\subsection{Method of analysis}

Since the near-UV through blue spectral regions comprising lines of CH, 
NH, and OH molecules are considerably crowded with lines, abundance 
determination should be done by comparing the observed and theoretically 
synthesized spectra with each other. For this purpose, Takeda's (1995) 
automatic spectrum-fitting technique was applied by employing the MPFIT 
program written by Y. Takeda based on Kurucz's (1993a) WIDTH9 code.
This method aims to match the observed (flux) spectrum in an arbitrary 
scale ($f^{\rm obs}_{\lambda}$) with the theoretical flux $F^{\rm th}_{\lambda}$,
which is a function of various parameters; e.g., wavelength 
shift ($\Delta\lambda$), macrobroadening velocity ($v_{\rm M}$: 
$e$-folding half-width of the Gaussian macrobroadening function, 
$f_{\rm M}(v) \propto \exp [(-v/v_{\rm M})^{2}]$), elemental abundances 
($A_{1}$, $A_{2}$, $\ldots$), etc.
In addition, two temporary parameters for adjusting 
$f^{\rm obs}_{\lambda}$ are also included: $C$ (scale control; cf. 
sect.~2 in Takeda 1995) and $\beta$ (tilt control; cf. footnote~3 
in Takeda \& Tajitsu 2014). The final solutions accomplishing the best 
match between theory and observation are obtained by numerically 
searching the multi-parameter space by applying the Newton--Raphson
algorithm. Note that this method does not require any necessity of
normalizing the observed spectrum by the local continuum in advance, 
which is a distinct merit because precisely placing the continuum 
level is very difficult in the present case. 

In order to fit the line features of XH molecules (X = C or N or O), 
a depth-independent scale factor $\phi$(XH) was introduced,
by which the number population of XH molecules (computed from a model 
atmosphere with the metallicity-scaled X abundance $A_{\rm model}$(X)) 
is to be multiplied to reproduce the observed  XH line strengths. 
Then, the X abundance of a star $A$(X) was assumed to relate with 
$\phi$(XH) as $A$(X) = $A_{\rm model}$(X) + $\log\phi$(XH).
Note that  this implicitly assumes that the population of XH
is proportional to the composition of X, which is practically valid
in the atmospheric condition of solar-analog stars where the number 
population of XH molecule is only a minor fraction of the total 
number of X nuclei (i.e., most are in the form of neutral atoms). 

\subsection{Spectral regions and line data}

The spectral regions comprising CH, NH, and OH lines, where 
the fitting analysis for abundance determination is performed,
were selected after exploratory test runs at 4270--4330~\AA\ (CH), 
3340--3390~\AA\ (NH), and 3100--3200~\AA\ (OH).
The finally adopted regions (12, 11, and 11 for CH, NH, and OH; each 
being typically $\sim$~0.5--2~\AA\ wide) are summarized in Table~2.
This table also contains the elements whose abundances were varied 
(along with $\Delta\lambda$ and $v_{\rm M}$) in the iterative 
analysis, while the abundances of all other elements were fixed at 
the metallicity-scaled solar abundances.

Regarding the molecular line data of CH, NH, and OH in these regions, 
the files ``ch.asc'', ``nh.asc'', and ``oh.asc'' downloaded from Kurucz's 
homepage (http://kurucz.harvard.edu/linelists/linesmol/) were invoked. 
The $\log gf$ values in these files were further scaled with the 
standard isotope ratios by using Kurucz's (1993b) ``RMOLEC.FOR'' program.
Meanwhile, the data of atomic lines were taken from the VALD database 
(Ryabchikova et al. 2015). In case that damping parameters are not 
available, the default treatments used in Kurucz's (1993a) WIDTH9
program were employed. The finally adopted line data are presented
as the supplementary electronic data (directory ``linedat''; 
cf. Appendix A).   

As to the dissociation potentials ($D_{0}$), the data already incorporated 
in the WIDTH9 code were adopted for CH (3.465~eV) and OH (4.392~eV) unchanged,
which are almost the same as used by Su\'{a}rez-Andr\'{e}s et al. (2017) 
and Ecuvillon et al. (2006), respectively. However, $D_{0}$ for NH was
replaced by 3.37~eV (instead of the original 3.47~eV) according to
Su\'{a}rez-Andr\'{e}s et al. (2016).

\setcounter{table}{1}
\begin{table*}[h]
\begin{center}
\caption{Adopted regions and varied abundances in the fitting analysis.}
\setlength{\tabcolsep}{5.0pt}
\small
\begin{tabular}{ccccl}
\hline\hline
No. & Region code & $\lambda_{1}$ & $\lambda_{2}$ & Varied abundances\\ 
\hline
\multicolumn{5}{c}{(C abundance determination)}\\
 1 & CH4273 & 4273.02 & 4274.36 & Fe, CH                  \\
 2 & CH4277 & 4276.49 & 4277.94 & Fe, CH, Zr              \\
 3 & CH4280 & 4279.19 & 4281.48 & Mn, Fe, CH, Cr, Ti, Sm  \\
 4 & CH4284 & 4283.90 & 4285.25 & CH, Ni, Cr, Fe, Ti, Mn  \\
 5 & CH4286 & 4285.62 & 4287.21 & Ti, Fe, CH              \\
 6 & CH4295 & 4295.54 & 4296.44 & Ti, Ni, Cr, CH          \\
 7 & CH4297 & 4297.37 & 4298.42 & Fe, CH, Cr              \\
 8 & CH4309 & 4309.23 & 4310.33 & Fe, Y, CH               \\
 9 & CH4312 & 4312.39 & 4313.31 & Ti, CH, Mn, Fe          \\
10 & CH4313 & 4313.30 & 4314.60 & Sc, Fe, CH              \\
11 & CH4323 & 4322.41 & 4324.57 & CH                      \\
12 & CH4328 & 4327.40 & 4328.81 & Fe, CH                  \\
\hline
\multicolumn{5}{c}{(N abundance determination)}\\
 1 & NH3340 & 3340.10 & 3341.04 & Ti, Fe, NH              \\
 2 & NH3357 & 3356.98 & 3357.63 & Zr, Cr, NH, Fe          \\
 3 & NH3358 & 3357.63 & 3358.38 & Ti, NH, Fe              \\
 4 & NH3362 & 3362.50 & 3363.07 & Ni, Ti, NH, Cr          \\
 5 & NH3363 & 3363.11 & 3363.86 & Cr, Ni, Fe, NH, Zr, Co  \\
 6 & NH3364 & 3364.45 & 3365.35 & Ni, NH, Fe, Co          \\
 7 & NH3365 & 3365.32 & 3366.04 & Ni, NH, Fe              \\
 8 & NH3370 & 3370.09 & 3371.22 & Fe, Ti, Co, NH          \\
 9 & NH3382 & 3382.10 & 3382.84 & Cr, Fe, Ti, NH          \\
10 & NH3387 & 3386.95 & 3387.70 & Fe, Ni, Co, NH          \\
11 & NH3388 & 3388.36 & 3389.07 & Ti, Fe, NH              \\
\hline
\multicolumn{5}{c}{(O abundance determination)}\\
 1 & OH3103 & 3102.79 & 3103.95 & Ti, Fe, Cr, OH          \\
 2 & OH3105 & 3105.25 & 3106.43 & Ti, Ni, Fe, OH          \\
 3 & OH3117 & 3117.51 & 3118.09 & Ti, OH, Fe              \\
 4 & OH3124 & 3123.62 & 3124.60 & OH, Ti, Fe              \\
 5 & OH3126 & 3125.83 & 3127.17 & V, Fe, Zr, OH           \\
 6 & OH3141 & 3141.36 & 3142.61 & Fe, V, OH, Ti           \\
 7 & OH3147 & 3146.83 & 3147.71 & Cr, Fe, Co, OH          \\
 8 & OH3149 & 3149.57 & 3150.18 & Cr, OH, Fe              \\
 9 & OH3168 & 3168.23 & 3169.27 & Ti, Fe, OH, Cr          \\
10 & OH3189 & 3188.69 & 3189.66 & Fe, OH, Ti              \\
11 & OH3191 & 3191.48 & 3192.14 & Fe, Ti, OH, Zr, Ni      \\
\hline
\end{tabular}
\end{center}
In Columns 3 and 4, $\lambda_{1}$ and $\lambda_{2}$ are
the starting and ending wavelengths (in \AA) of the spectral region
where the fitting analysis was done.  
\end{table*}

\subsection{Analysis results}

The iterative solution converged successfully in most of the 34 (=12+11+11) 
regions for the 118 stars as well as the Sun (Vesta), though some parameters 
(abundances or $v_{\rm M}$) had to be fixed in exceptional cases (especially 
for broader-line stars of comparatively higher rotational velocity) in order 
to avoid instability or divergence.

How the theoretical spectrum for the converged parameter solutions match
the observed one is illustrated for each of the spectral regions in Figure~1 (CH),
Figure~2 (NH), and Figure~3 (OH) for the case of the Sun.
(The information regarding which lines of which species contribute to 
the complex spectral features of these figures may be found from the line 
data files in the directory ``linedat'' mentioned in Sect.~3.2; i.e., 
line-to-continuum opacity ratios $\eta$ computed for all lines are useful.)     
Similar figures showing the accomplished spectrum fit for all 118 stars 
and the relevant data of observed/theoretical spectra  
are presented as the supplementary electronic data (directories ``fitfigs'' 
and ``specdat''; cf. Appendix A).

The resulting CNO abundances for the Sun [$A_{\odot}$(C), $A_{\odot}$(N), 
and $A_{\odot}$(O)] derived for each region are displayed together in Figure~4, 
where the reference solar abundances taken from two compilations 
(Anders \& Grevesse 1989; Asplund et al. 2009) are also shown for comparison.
We can see from this figure that CH, NH, and OH lines in the blue or near-UV 
region tend to yield more or less lower abundances (especially for N and O).
as compared to the actual values. This is the tendency already found by
Laird (1985) (cf. Sect.~1). The reason for this systematic error
is not clear, for which several possibilities may be considered; such as
missing opacity, overdissociation, 3D effect, etc.
In any event, this problem involving the absolute scale of $A$ is irrelevant 
in the present case, because we aim to do a purely differential region-by-region 
analysis relative to the Sun.   

\begin{figure}[h] 
\begin{minipage}{70mm}
\begin{center}
   \includegraphics[width=7.0cm]{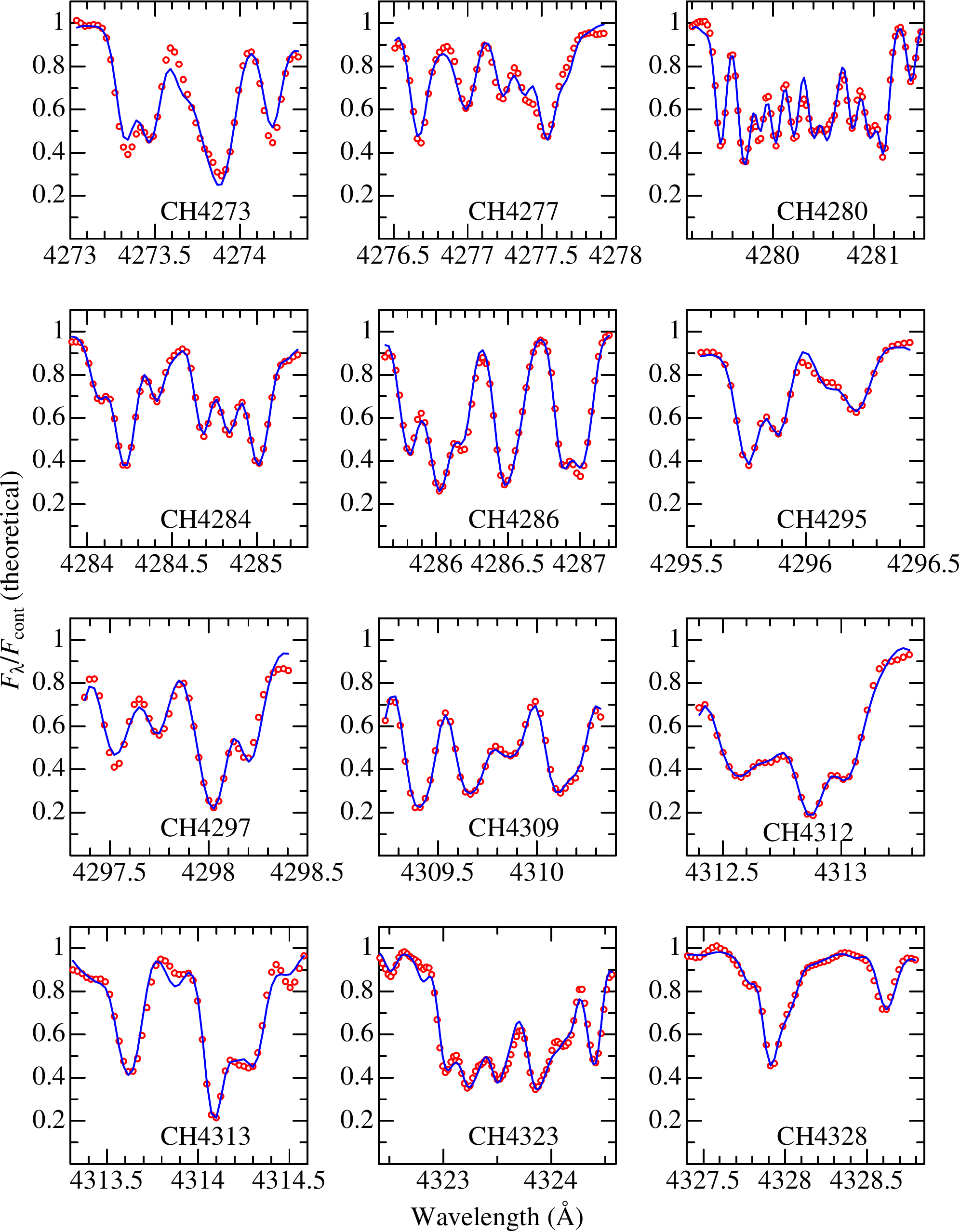}
\end{center}
\caption{Observed and fitted theoretical spectra in each of the 12 regions 
(within 4270--4330~\AA), where C abundances were determined from CH lines. 
Shown here is the representative case of the solar (Vesta) spectrum.
The observed and theoretical spectra are depicted by red open symbols 
and blue lines, respectively. The corresponding region code (cf. Table~2) 
is specified in each panel. The wavelength scale of the spectrum is 
adjusted to the laboratory frame, and the scale marked in the left ordinate 
corresponds to the theoretical residual flux 
($F^{\rm th}_{\lambda}/F^{\rm th}_{\rm cont}$).
} 
   \label{Fig1}
\end{minipage}
\end{figure}

\begin{figure}[h]
\begin{minipage}{70mm}
\begin{center}
  \includegraphics[width=7.0cm]{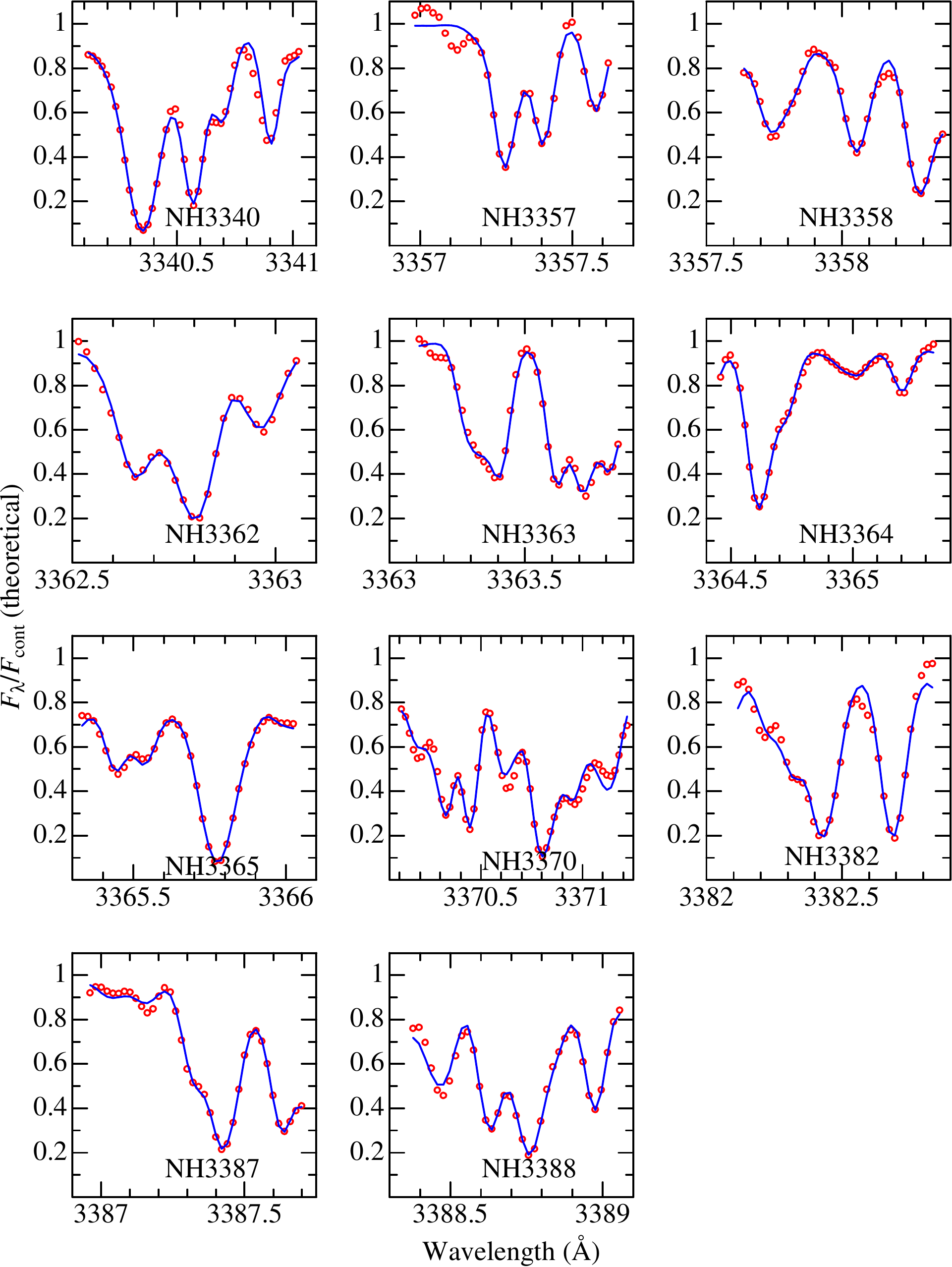}
\end{center}
   \caption{
Observed and fitted theoretical solar spectra in each of the 
11 regions (within 3340--3390~\AA), where N abundances were determined 
from NH lines. Otherwise, the same as in Figure~1.
} 
   \label{Fig2}
\end{minipage}
\end{figure}

\begin{figure}[h]
\begin{minipage}{70mm}
\begin{center}
  \includegraphics[width=7.0cm]{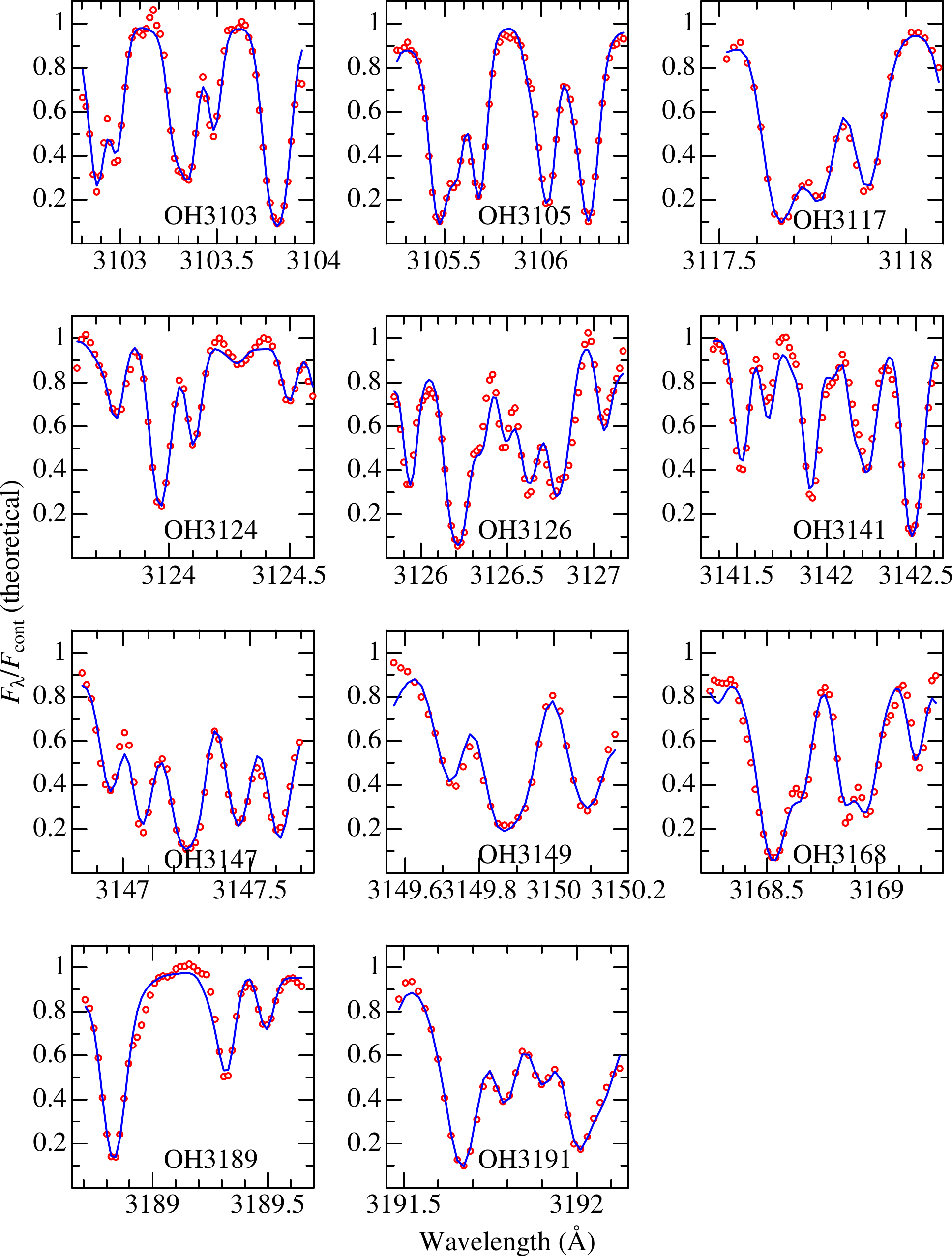}
\end{center}
   \caption{
Observed and fitted theoretical solar spectra in each of the 
11 regions (within 3100--3200~\AA), where O abundances were determined 
from OH lines. Otherwise, the same as in Figure~1.
} 
   \label{Fig3}
\end{minipage}
\end{figure}

\begin{figure}[h] 
\begin{minipage}{70mm}
\begin{center}
   \includegraphics[width=7.0cm]{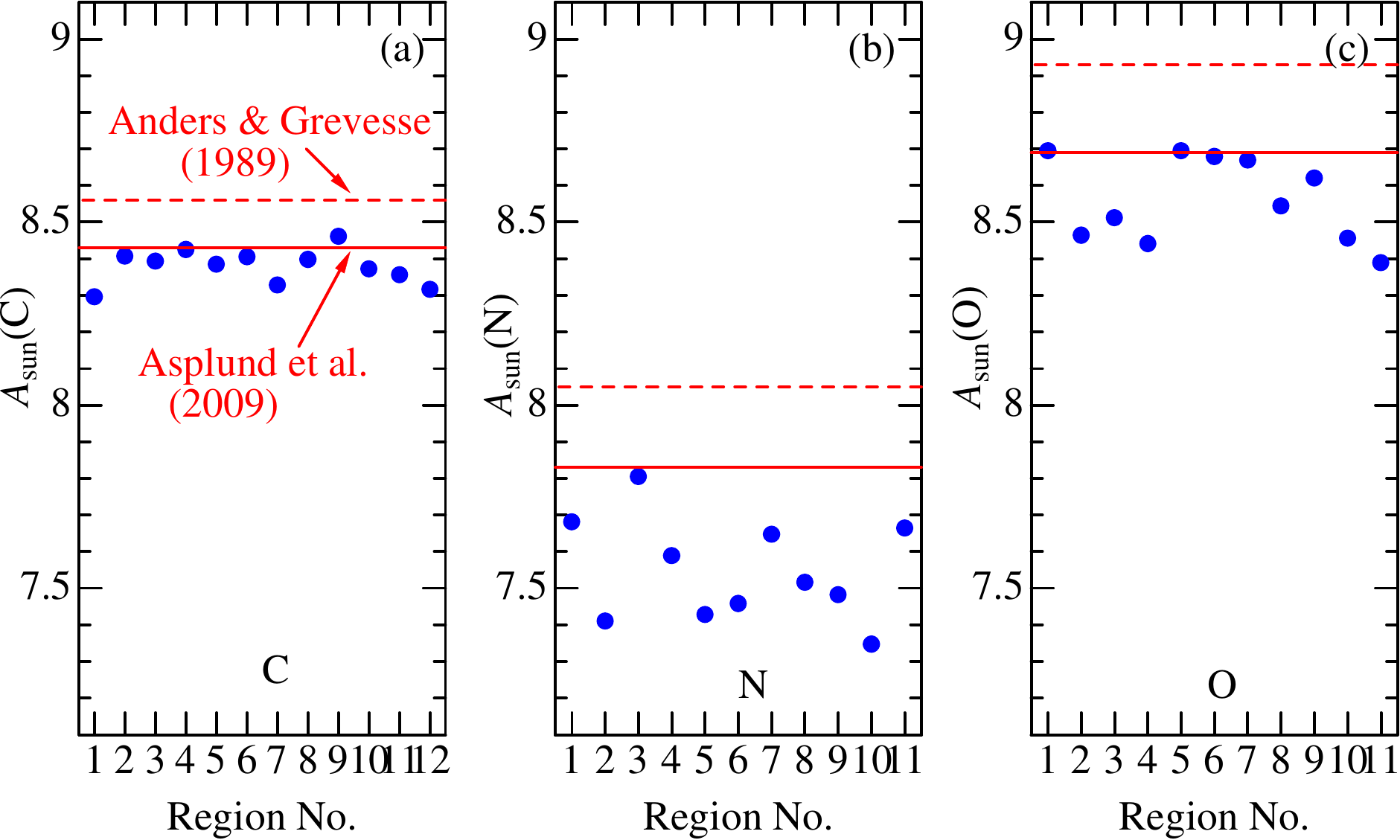}
\end{center}
\caption{
The results of C, N, and O abundances for the Sun ($A_{\odot}$; logarithmic 
number abundance in the usual normalization of $A_{\rm H} = 12$), which were 
derived from the fitting analysis of solar (Vesta) spectrum in each region, 
are plotted against the corresponding region number. The left, middle, and 
right panels are for C, N, and O, respectively.
The reference solar CNO abundances published by Anders \& Grevesse (1989) 
(8.56, 8.05, 8.93) and Asplund et al. (2009) (8.43, 7.83, 8.69) are indicated 
by the horizontal dashed and solid lines, respectively.
} 
   \label{Fig4}
\end{minipage}
\end{figure}

\subsection{Mean abundances and their errors}

Let the abundance of X (= C or N or O) derived from region $i$ be $A_{* i}$ (star) or 
$A_{\odot i}$ (Sun). Then, the mean differential abundance relative to the Sun 
averaged over the available regions is defined as
\begin{equation}
\langle[{\rm X/H}]\rangle \equiv \sum_{i=1}^{n_{\rm s}}{([{\rm X/H}]_{i})} / n_{\rm s}
\equiv \sum_{i=1}^{n_{\rm s}}{(A_{* i} - A_{\odot i})}/n_{\rm s}  \; , 
\end{equation}
where $n_{\rm s}$ is the number of selected regions finally used for averaging.\footnote{
This number may be equal to or smaller than the total number of regions ($n_{\rm t}$ = 12, 11, 
and 11 for C, N, and O), because outlier $[{\rm X/H}]_{i}$ values judged by Chauvenet's criterion
were rejected.}
Since the standard deviation around this mean is
\begin{equation}
\sigma \equiv 
\sqrt{\sum_{i=1}^{n_{\rm s}}{([{\rm X/H}]_{i} - \langle[{\rm X/H}]\rangle)^{2}}/n_{\rm s}} \; ,
\end{equation} 
the mean error involved in $\langle[{\rm X/H}]\rangle$ is written as
\begin{equation}
\epsilon \equiv \sigma/\sqrt{n_{\rm s}} \; .
\end{equation} 
Figure~5 (upper 9 panels) shows the dispersion of each $[{\rm X/H}]_{i}$ around 
$\langle[{\rm X/H}]\rangle$ and the extent of mean error ($\pm\epsilon$) 
for all 118 stars, while the distribution histograms of $\epsilon$ are illustrated
in the bottom 3 panels. As seen from these histograms, most $\epsilon$ values 
are within $\la$~0.01--0.02~dex, although several stars (mostly those of broader lines)
exceptionally show larger $\epsilon$ amounting up to $\sim$~0.03--0.04~dex.
Actually, mean $\epsilon$ values averaged all stars are ($\langle \epsilon_{\rm C} \rangle$,
$\langle \epsilon_{\rm N} \rangle$, $\langle \epsilon_{\rm O} \rangle$) = 
(0.007, 0.011, 0.011)~dex.
The detailed results of each region's [X/H]$_{i}$, $\langle[{\rm X/H}]\rangle$,
$\sigma$, and $\epsilon$ (X = C, N, O) for all the program stars are
summarized in the files ``relabs\_ch.dat'', ``relabs\_nh.dat'', and ``relabs\_oh.dat''
(placed in the directory ``abunds'') of the supplementary material (cf. Appendix A).

Another error source we have to take into consideration is ambiguities in the 
atmospheric parameters. As mentioned in Sect.~2.2, the typical statistical errors 
involved in $T_{\rm eff}$, $\log g$, and $v_{\rm t}$ are $\sim \pm 20$~K, 
$\sim \pm 0.05$~dex, and $\sim \pm 0.1$~km~s$^{-1}$,respectively. 
In order to estimate the impact of these errors, the fitting analysis for the 
solar spectrum was repeated by perturbing these parameters interchangeably 
to see the resulting abundance changes ($\delta_{T\pm}$, $\delta_{g\pm}$, $\delta_{v\pm}$) 
and their root-mean-square ($\delta_{Tgv}$).
The results are depicted in Figure~6, which indicates that $\delta_{Tgv}$ is
essentially determined by $\delta_{T\pm}$ (reflecting the large $T$-sensitivity) 
and $\delta_{Tgv}$ is typically $\sim$~0.02--0.03~dex.
As this $\delta_{Tgv}$ acts rather similarly to each [X/H]$_{i}$ (i.e., not random 
but in the same direction), $\langle[{\rm X/H}]\rangle$ suffers also this amount 
of ambiguities due to parameter uncertainties (maily determined by that of $T$).    

Accordingly, combining these two kinds of errors ($\epsilon$ and 
$\delta_{Tgv}$), the typical extent of total error involved in 
$\langle[{\rm X/H}]\rangle$ would finally make $\la 0.03$~dex.

\begin{figure}[h]
\begin{minipage}{70mm}
\begin{center}
  \includegraphics[width=7.0cm]{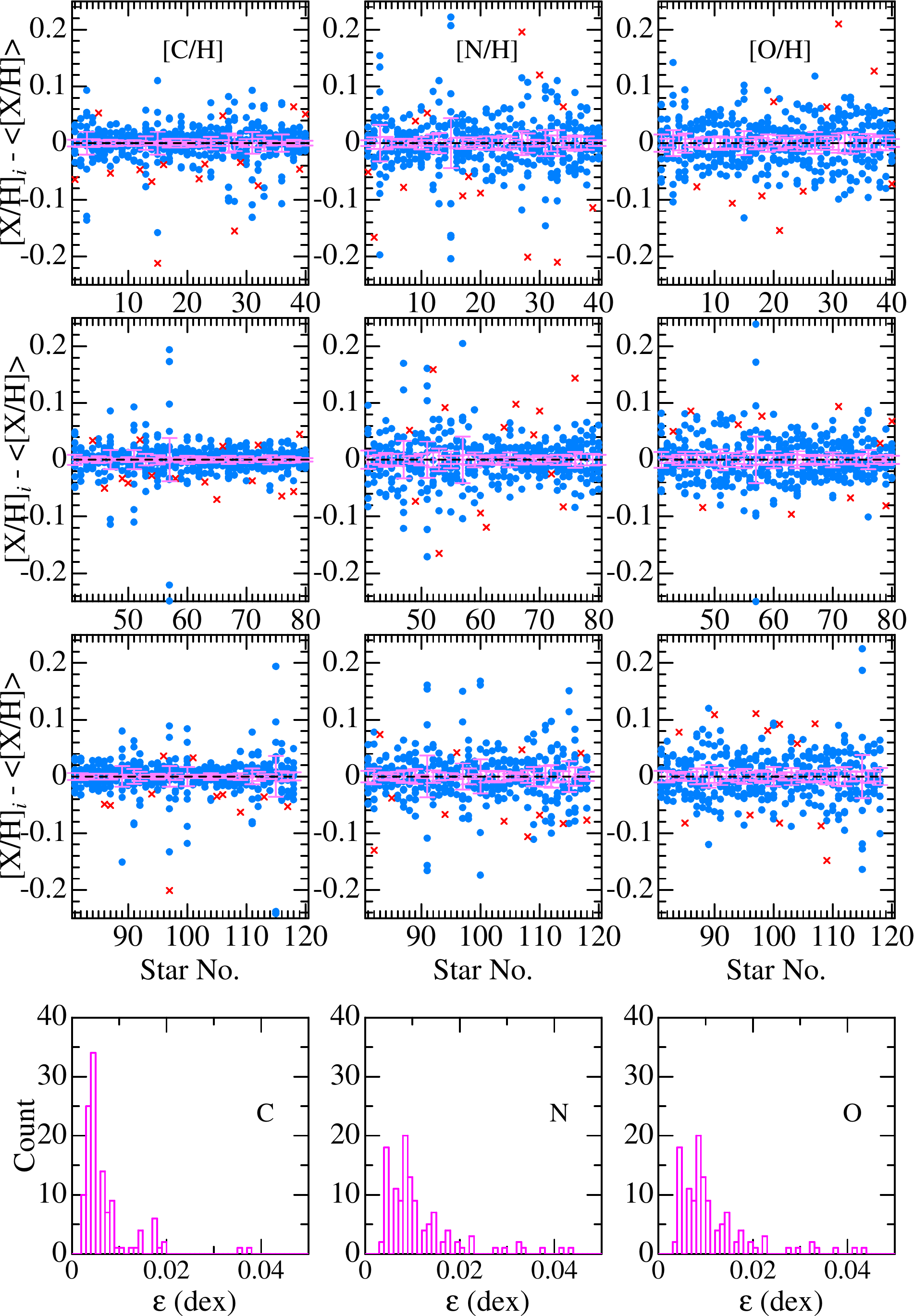}
\end{center}
   \caption{
In the upper 9 panels, differences between each [X/H] (X is C/N/O
in the left/middle/right panels) of region $i$ ([X/H]$_{i}$) and the 
mean [X/H] over the regions ($\langle$[X/H]$\rangle$) are plotted 
(in blue symbols) for each of the 118 program stars, which helps to 
understand the extent of dispersion around the mean. The outlier values 
(judged by Chauvenet's criterion) excluded from the averaging process 
are indicated by red crosses. The pink error bars show the size of 
mean error ($\pm \epsilon$) of $\langle$[X/H]$\rangle$.
In the lowest three panels are shown the distribution histograms for 
$\epsilon$.
} 
   \label{Fig5}
\end{minipage}
\end{figure}

\begin{figure}[h]
\begin{minipage}{70mm}
\begin{center}
  \includegraphics[width=7.0cm]{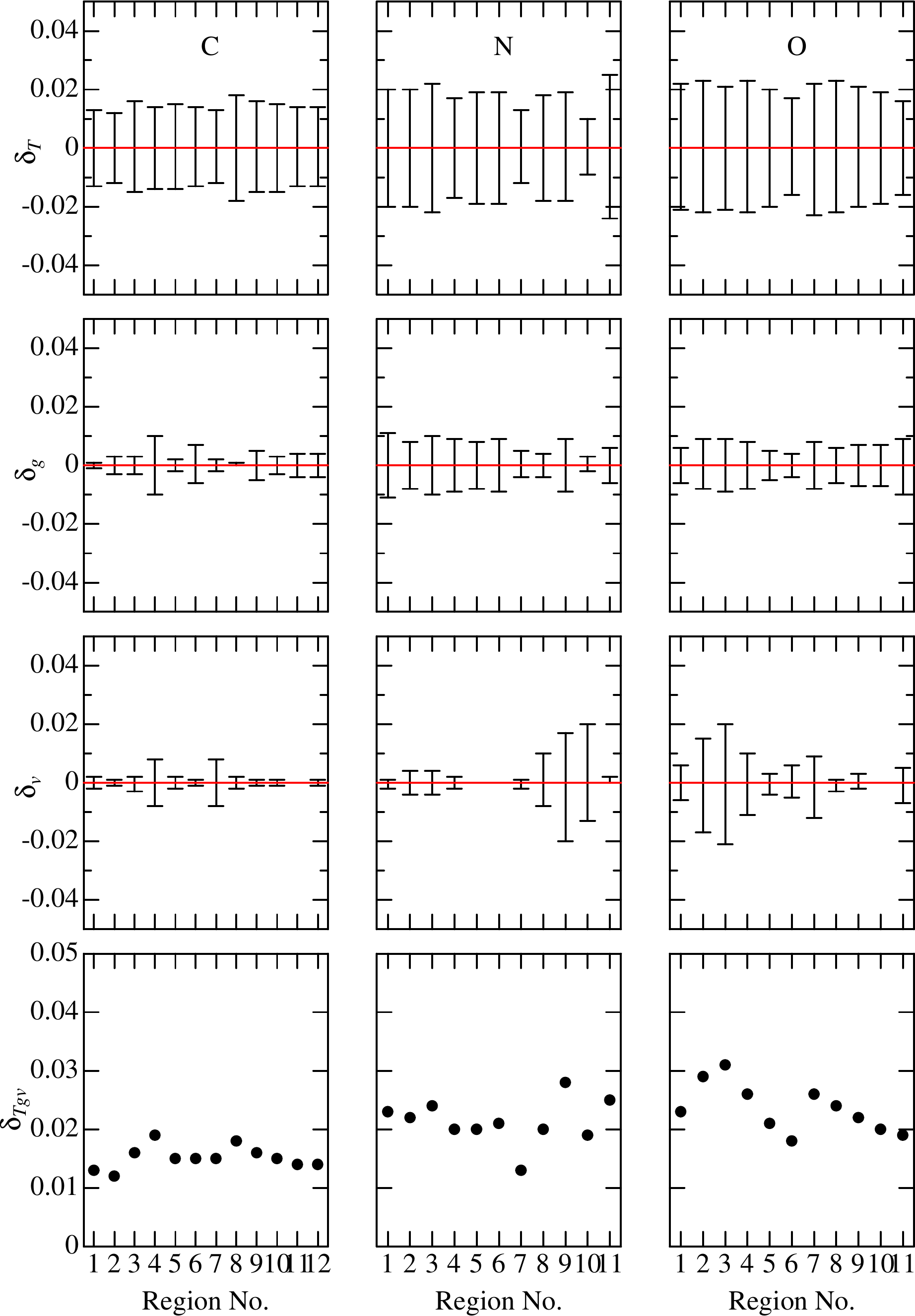}
\end{center}
   \caption{
Effect on changing atmospheric parameters on the [X/H] result
of each region calculated for the representative case of the Sun, 
where left, middle, and right panels are for C, N, and O, respectively. 
1st-row panels: $\delta_{T+}$ and $\delta_{T-}$ (abundance variations 
in response to $T_{\rm eff}$ changes by $\pm 20$~K). 
2nd-row panels: $\delta_{g+}$ and $\delta_{g-}$ (abundance variations 
in response to $\log g$ changes by $\pm 0.05$~dex). 
3rd-row panels: $\delta_{v+}$ and $\delta_{v-}$ (abundance 
variations in response to perturbing the $v_{\rm t}$ value
by $\pm 0.1$~km~s$^{-1}$).
4th-row panels: root-sum-square of these three kinds of $\delta$'s, which is defined as 
$\delta_{Tgv} \equiv (\delta_{T}^{2} + \delta_{g}^{2} + \delta_{v}^{2})^{1/2}$,
where  $\delta_{T} \equiv (|\delta_{T+}| + |\delta_{T-}|)/2$, 
$\delta_{g} \equiv (|\delta_{g+}| + |\delta_{g-}|)/2$, and
$\delta_{v} \equiv (|\delta_{v+}| + |\delta_{v-}|)/2$.
} 
   \label{Fig6}
\end{minipage}
\end{figure}

\section{Discussion}

\subsection{CNO-to-Fe abundance ratios}

Now that the relative abundances [X/H]\footnote{
Hereinafter, the notation [X/H] is used to indicate $\langle$[X/H]$\rangle$
(mean value averaged over the regions) for simplicity.} 
(X = C, N, O) have been established, we can examine the trends of [X/Fe] 
($\equiv$~[X/H]$-$[Fe/H]; logarithmic X-to-Fe abundance ratio).
derived for solar-analog stars from hydride molecules in comparison with
previous results, especially with those of Takeda \& Honda (2005) 
determined for solar-type stars in a broader sense (FGK dwarfs/subgiants) 
by using atomic lines.
How the resulting [X/H] and [X/Fe] are correlated with [Fe/H] (and also
the mutual correlation between C and O) is displayed in the left and middle 
panels of Figure~7. The mean values averaged over each 0.1~dex bin of [Fe/H] 
between $-0.3\le$~[Fe/H]~$\le +0.3$ are illustrated in the right panels 
of this figure, where the corresponding Takeda \& Honda's (2005)
results for FGK stars (derived from C~{\sc i} 5052/5380, N~{\sc i} 8680, and 
O~{\sc i} 7771--5 lines shown in their fig.~6) are also plotted for comparison.  

We can see that [C/H] almost scales with [Fe/H] (Fig.~7a), while a systematic 
departure begins to appear for [N/H] ($<$~[Fe/H]; Fig.~7b) and [O/H] ($>$~[Fe/H]; 
Fig.~7c) with a decrease in [Fe/H]. As a result, although [C/Fe] does not show 
so clear [Fe/H]-dependence (Fig.~7e), [N/Fe] tends to decrease (Fig.~7f) 
while [O/Fe] increases (Fig.~7g) with a lowering of [Fe/H]. Accordingly,
[C/O] (= [C/Fe]$-$[O/Fe]) exhibits a progressive decrease towards lower metallicity. 
(Fig.~7f). The linear-regression relations determined by the least-squares fit are
[C/Fe] = $-0.046 (\pm 0.032)$[Fe/H]~$- 0.067 (\pm 0.005)$,
[N/Fe] = $+0.242 (\pm 0.037)$[Fe/H]~$- 0.068 (\pm 0.006)$,
[O/Fe] = $-0.350 (\pm 0.027)$[Fe/H]~$- 0.035 (\pm 0.004)$, and 
[C/O]  = $+0.305 (\pm 0.025)$[Fe/H]~$- 0.031 (\pm 0.004)$,
as depicted also in the relevant figure panels.

The slopes (d[X/Fe]/d[Fe/H]) of these relations are important  
in connection with the galactic chemical evolution. Yet, care should be taken  
in comparing them with other work, because the resulting gradient may depend 
upon how the sample stars are chosen. Particularly, since our solar-analog stars 
cover only a comparatively narrow metallicity range (most stars are within 
$-0.3 \la$~[Fe/H]~$\la +0.3$), they are rather disadvantageous in this respect.
Keeping this in mind, we may state that these trends are more or less in accord 
(at least qualitatively) with those published in previous studies; e.g., 
Takeda \& Honda (2005) for C and O (note that their results for N suffer large 
uncertainties and thus unreliable), Ecuvillon et al. (2006) for O,  
Su\'{a}rez-Andr\'{e}s et al. (2016) for N,
Su\'{a}rez-Andr\'{e}s et al. (2017) for C,
Delgado Mena et al. (2021) for C/O,
and the references therein.

Yet, some differences are noticeable from a quantitative point of view.
Especially, the resulting d[C/Fe]/d[Fe/H] slope of $-0.05$ is apparently shallower 
than the previous values (e.g., $\sim -0.2$ concluded by Takeda \& Honda 2005); 
but this is due to the fact that (i) The [Fe/H]-dependence of [C/Fe] not linear 
but shows an appreciable upturn around [Fe/H]$\sim 0$ and (ii) the [Fe/H] range of 
our sample stars is narrow ($\pm$ several tenths dex around the solar metallicity).
For this reason, the gradient for [C/Fe] (and [C/O]) derived here should not 
be seriously taken.

Meanwhile, attention should be paid also to the intercept values of these regression 
relations at [Fe/H]~$=0$ ($-0.067$, $-0.068$, and $-0.035$~dex for C, N, and O). 
That is, the gravity center in the distributions of [C/Fe], [N/Fe], and [O/Fe] 
ratios around [Fe/H]~$\sim 0$ is not zero but slightly negative by several 
hundredths dex, which can also be recognized by eye-inspection of Figure~7e, 
7f, and 7g (or from Fig.~7i, 7j, and 7k; blue bullet symbols).
Since such a shift was not found in Takeda \& Honda's (2005) results for 
FGK stars (cf. pink bullets in Fig.~7i, 7j, and 7k), this detection
is a consequence of high-precision relative abundances, which could be 
accomplished thanks to the effective differential analysis between the 
Sun and solar-analog stars. This zero-point offset in [C/Fe], [N/Fe], 
and [O/Fe] is a significant feature in relation to the status of our Sun 
among the solar-analog stars, as mentioned in the next Section~4.2.

\begin{figure*}[h]
\begin{minipage}{160mm}
\begin{center}
  \includegraphics[width=12.0cm]{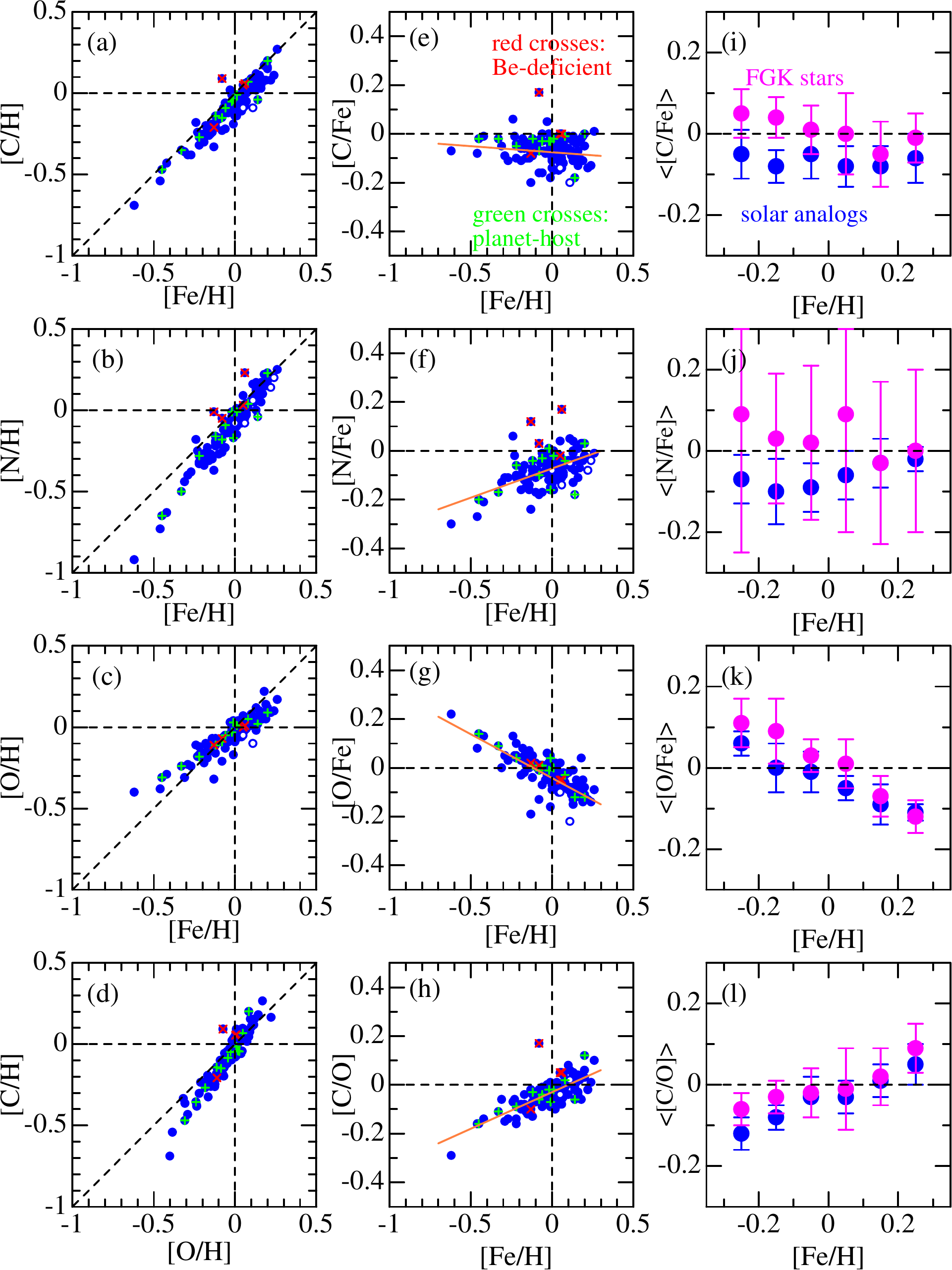}
\end{center}
   \caption{
In the left- and middle-column panels are shown the correlations
between [X/H] (logarithmic abundance of element X relative to the Sun;
X = C or N or O or Fe) and their differences (e.g., 
[C/Fe] $\equiv$ [C/H] $-$ [Fe/H]): 
(a) [C/H] vs. [Fe/H], (b) [N/H] vs. [Fe/H], (c) [O/H] vs. [Fe/H], 
(d) [C/H] vs. [O/H]. (e) [C/Fe] vs. [Fe/H], (f) [N/Fe] vs. [Fe/H],
(g) [O/Fe] vs. [Fe/H], and (h) [C/O] vs. [Fe/H].
Less reliable values ($\epsilon > 0.03$~dex) are indicated by open symbols; 
and green crosses (+) and red crosses ($\times$) are overplotted for 
12 planet-host stars and 4 Be-deficient stars, respectively.
In panels (e)--(h), the linear-regression lines determined by the least-squares 
analysis are also drawn by the orange solid lines.
The bullet symbols in the right-column panels are the mean $\langle$[X/Y]$\rangle$ 
at each metallicity group (0.1~dex bin within $-0.3 \le$~[Fe/H]~$\le +0.3$) 
determined for 118 solar analogs (blue, this study) and 160 FGK dwarfs (pink, 
Takeda \& Honda 2005; the results shown in their fig.~6) for comparison, 
where error bars denote the standard deviations ($\sigma$). Panels (i), (j), 
(k), and (l) are for $\langle$[C/Fe]$\rangle$, $\langle$[O/Fe]$\rangle$, 
$\langle$[N/Fe]$\rangle$, and $\langle$[C/O]$\rangle$, respectively.
} 
   \label{Fig7}
\end{minipage}
\end{figure*}

\subsection{Star--planet connection}

In order to examine whether stars harboring giant planets show any difference 
in their CNO abundances, 12 planet-host stars included in our 118 program 
stars are discriminated by overplotting green crosses in Figure~7a--7h.
These figures do not reveal distinct differences between planet-host 
stars and no-planet samples (i.e., most of the green crosses distribute 
near to the red linear-regression line showing the mean trend). 
Yet, some systematic tendency of [X/Fe](with planets) being slightly higher 
than [X/Fe](without planets) might be seen (especially for [C/Fe] in Fig.~7e).
In order to check this point quantitatively, the difference ($d$) between 
[X/Fe] for each planet-harboring star and $\langle$[X/Fe]$\rangle$ 
(mean value in the relevant metallicity bin) was calculated and compared with 
$\sigma$ (standard deviation of the distribution at the coresponding bin), 
as done by Takeda \& Honda (2005) (cf. sect.~5.2 and table~3 in their paper). 
It then turned out that, while $\sim$70\% of the $d$ values of these stars 
are positive (i.e., relatively overabundant trend on the average), the deviations 
($d$) do not exceed $1.5\sigma$ in most cases ($d \sim 1.7$ at the largest). 
This makes us feel that it is still premature to consider this trend as real.
Further studies on a much larger sample of planet-host stars would be required
to confirm or disconfirm the reality of this suspected tendency.  
Accordingly, a conservative statement is retained for the time being that the 
photospheric CNO abundances of solar-analog stars are not significantly affected 
by whether they host giant planets or not. This conclusion is almost
consistent with the consequences of Su\'{a}rez-Andr\'{e}s et al. (2016) for N, 
and Su\'{a}rez-Andr\'{e}s et al. (2017) for C. However, it does not lend 
support to Ecuvillon et al.'s (2006) argument for O that planet-host 
stars appear to show an oxygen overabundance by $\sim$~0.1--0.2~dex 
in comparison with the reference sample.

As a topic relevant to the influence of planet formation upon photospheric 
abundances of a host star, the issue of zero-point offset in the [X/Fe] 
distribution at [Fe/H]~$\sim 0$ (cf. Sect.~4.1) has to be mentioned again.
In order to ascertain the results described there, the [X/Fe] values in 
the near-solar metallicity range of $-0.1 \le$~[Fe/H]~$\le +0.1$ (comprising 
62 stars) were averaged, and the mean values 
($\langle$[X/Fe]$\rangle$) along with their mean errors 
($\pm \epsilon$) turned out  
$\langle$[C/Fe]$\rangle = -0.063 (\pm 0.007)$,
$\langle$[N/Fe]$\rangle = -0.070 (\pm 0.008)$, and 
$\langle$[O/Fe]$\rangle = -0.033 (\pm 0.005)$, which are in agreement
with the [Fe/H]~=~0 intercepts resulting from the linear-regression 
analysis in Section~4.1. Therefore, it is certain that these mean 
$\langle$[X/Fe]$\rangle$ values are slightly negative (by $\simeq$~0.06--0.07~dex 
for C and N, $\simeq$~0.03~dex for O) at [Fe/H]~$\sim 0$.

As a matter of fact, this is closely related to 
the finding of Mel\'{e}ndez et al. (2009), who reported based on
the high-precision differential analysis of 11 solar twins relative 
to the Sun that the Sun shows a characteristic chemical signature
in comparison with the reference sample of solar twins that 
the refractory elements (such as Fe) are comparatively deficient 
relative to the volatile ones (such as CNO) by $\sim 20$\%,
which may be associated with the formation mechanism of our solar system 
(especially rocky terrestrial planets).
As seen from their figure~2, when compared at the same solar metallicity,
the volatile elements (CNO) in the Sun are comparatively overabundant 
than the average of solar twins by $\simeq$~0.05~dex (C),
$\simeq$~0.06~dex (N), and $\simeq$~0.03~dex (O), which implies 
that the mean $\langle$[X/Fe]$\rangle$ of the reference stars at 
[Fe/H]~$\sim 0$ would turn out negative by these amounts. These 
offset values are satisfactorily consistent with our results.
Accordingly, our analysis of 118 solar analogs based on the lines of 
CH, NH, and OH yielded essentially the same conclusion as they obtained 
from 11 solar twins (although the lines used for abundance determination 
are not explicitly described in their paper, C~{\sc i}, N~{\sc i}, 
and O~{\sc i} lines are likely to have been invoked as seen from 
the wavelength range of their spectra).  

\subsection{CNO abundances of Be-dearth stars}

Takeda et al. (2011) reported that 4 stars out of 118 solar analogs
(program stars of this study) are strikingly Be-depleted (by $\ga 2$~dex). 
Actually, the lines of Be (and Li) are too weak and undetectable in these 
extraordinary stars (HIP~17336, 32673, 64150, and 75676).  
Soon after, Viallet \& Baraffe (2012)  investigated the impacts 
of rapid rotation and/or episodic accretion in the pre-main sequence phase 
(both may induce a global mixing, by which Li and Be are brought to the 
hot interior and burned out at temperatures of more than several million K) 
as a possible cause for such an extreme Be depletion.

From a different point of view, Desidera et al. (2016) pointed out 
that all these 4 peculiar solar analogs are binaries, and 
at least two of them (HIP~64150 and 75676) have white dwarf companions. 
This means that they may have suffered accretion of the nuclear-processed 
(Be-depleted) material from the evolved companion due to mass transfer 
events in the phase of red giant or asymptotic giant branch, which might 
be responsible for the Be anomaly.
This thought lead Desidera et al. (2016) to study the chemical 
abundances of C and s-process elements for these 4 stars in order to 
search for any signature of mass accretion from the companion. 
Interestingly, they found that HIP~75676 is an apparent barium star showing 
overabundances of s-process elements (Y, Zr, Ba, La) and C. Therefore, 
in order to supplement their investigation, it is worthwhile to check the CNO 
abundances we have determined for these Be-depleted stars (which are marked
 by red crosses in Fig.~7a--7h). The following consequences can be drawn.
\begin{itemize}
\item
The [C/Fe] values derived by Desidera et al. (2016) from the CH band at
4300~\AA\ ($-0.12$, $-0.05$, $0.00$, and $+0.21$ for HIP~17336, 32673, 
64150, and 75676, respectively) are in reasonable agreement with our 
results ($-0.08$, $0.00$, $0.00$, and $+0.17$).
\item
Appreciable anomalies deviating from the mean trend are seen in 
three cases for C and N (Fig.~7e and 7f): HIP~75675's [C/Fe] ($=+0.17$), 
HIP~17336's [N/Fe] ($=+0.12$), and HIP~32673's [N/Fe] ($=+0.17$).
In contrast, no peculiarity is seen in [O/Fe] for all 4 stars (Fig.~7g).
To sum up, three Be-depleted solar analogs (HIP~17336, 32673, and 75676) 
show some kind of anomaly in either C or N, while HIP~64150 is quite normal 
in terms of CNO abundances.
\item
As such, no conclusive evidence could be found for the tentative theory that 
considerable Be-depletion is caused by contamination of nuclear-processed 
materials from the companion. Although such an interaction event may have 
actually occurred in the past (especially for the barium star HIP~75676), 
the existence of Be-deficient CNO-normal star (HIP~64150) indicates 
that the solution to this problem is not so simple.
Besides, as Desidera et al. (2016) pointed out, even if such an efficient 
mass transfer takes place in the binary system, it is quantitatively
difficult to produce such a drastic Be depletion. 
\item
Accordingly, the question for the mechanism of depleting Be is still open.
The role of mass transfer from the companion might have an indirect effect 
on the deficiency of Be (e.g., induced thermohaline mixing or enhanced internal 
mixing triggered by episodic accretion), as discussed by Desidera et al. (2016).
Also, we should pay attention also to the possibility of Be-depletion caused 
by an effective mixing (e.g., due to rapid rotation) in the pre main-sequence 
phase, as discussed by Viallet \& Baraffe (2012). 
\end{itemize}  

\section{Summary and conclusion}

Clarifying the behaviors of C, N, and O abundances (representative light 
elements processed in the stellar core to be dredged up and ejected outwards
in the course of stellar evolution) in solar-type low-mass stars of diversified 
ages is important for studying the chemical evolution history of the Galaxy.

However, precisely establishing the key quantities [C/Fe], [N/Fe], and [O/Fe] 
(in comparison with the metallicity [Fe/H]) is not necessarily easy, because 
often adopted atomic C~{\sc i}, N~{\sc i}, and O~{\sc i} lines are small in 
number and generally weak. A possibility to ameliorate this situation is to
invoke the lines of hydride molecules (CH, NH, and OH) numerously available
with sufficiently large strengths in blue or near-UV wavelength regions.  
Although absolute abundances derived from these molecular lines are apt to 
suffer systematic errors, this problem can be circumvented by carrying out 
differential analysis relative to the Sun while limiting the sample only 
to solar-analog stars (early G-type dwarfs). 

This consideration motivated the author to determine the C, N, and O abundances 
of 118 solar-analog stars, whose atmospheric parameters ($T_{\rm eff}$, $\log g$,
$v_{\rm t}$, and [Fe/H]) are already established by Takeda et al. (2007), 
based on the lines of hydride molecules in blue or near-UV regions. 
For this purpose, extensive spectrum-synthesis analyses based on the efficient 
automatic fitting algorithm were applied to 12 spectral regions of CH lines 
(selected from 4270--4330~\AA),  11 regions of NH lines (from 3340--3390~\AA), 
and 11 regions of OH lines (from 3100--3200~\AA).

The primary aims of this study were (i) to clarify the behaviors of [C/Fe], 
[N/Fe], and [O/Fe], (ii) to examine whether any abundance characteristics 
related to the existence of planets is seen, and (iii) to check whether any 
anomaly exists in the CNO abundances of 4 drastically Be-depleted stars 
found by Takeda et al. (2011). 

The trends of [C/Fe], [N/Fe], and [O/Fe] in relation to [Fe/H] turned out
almost consistent (at least qualitatively) with those reported by past studies 
mainly based on atomic lines: In the metallicity range of $-0.6 \la$~[Fe/H]~$\la +0.3$, 
[C/Fe] shows a marginally increasing tendency for decrease of [Fe/H] with 
a slight upturn around [Fe/H]~$\sim 0$, [N/Fe] tends to somewhat
decrease towards lower [Fe/H], and [O/Fe] systematically increases (and thus 
[C/O] decreases) with decreasing [Fe/H].

It is noteworthy, however, that the gravity centers of these [X/Fe] ratios (X = C, N, O) 
are slightly subsolar (negative) by several hundredths dex ($\simeq$~0.06--0.07~dex
for C and N, $\simeq$~0.03~dex for O) around [Fe/H]~$\sim 0$, which may be 
interpreted as unusual CNO-to-Fe abundance ratios of the Sun (compared to the mean 
of other solar analogs). This is essentially a reconfirmation of the finding of 
Mel\'{e}ndez et al. (2009), who reported that refractory elements (such as Fe) are 
somewhat deficient relative to the volatile ones (such as CNO) in the solar photosphere 
in comparison with the sample of 11 solar twins, which they suspected may be related to 
the formation mechanism of our solar system (especially rocky terrestrial planets).

In the meanwhile, regarding the question whether CNO abundances suffer any influence 
by the existence of giant planets, clear differences are not seen in the 
distributions of [C/Fe], [N/Fe], and [O/Fe] for 12 planet-host stars in comparison 
to other no-planet samples, though a possibility of the former tending to be slightly 
larger than the latter can not be ruled out. 

Given that Desidera et al. (2016) reported that all 4 Be-depleted stars 
(HIP~17336, 32673, 64150, and 75676) detected by Takeda et al. (2011) are binary systems 
(especially at least 2 stars have white dwarf companions), it is worthwhile to examine
whether they have any CNO anomalies caused by contamination of nuclear-processed 
materials. Our results indicate that three of them (HIP~17336, 32673, and 75676) 
show overabundances in either C or N, whereas HIP~64150 is quite normal 
in terms of CNO abundances. As such, mass transfer from the companion may 
have actually occurred in these stars (especially, highly probable for the barium star 
HD~75676). However, it is premature to relate this to the cause of Be anomaly, 
because this mechanism alone is quantitatively difficult to produce such a 
drastic Be depletion (by $\ga 2$~dex).
Accordingly, the question for the mechanism of depleting Be is still open.
Several other interpretations such as those related to pre-main sequence 
evolution (Viallet \& Baraffe 2012) or indirect effect of mass transfer from 
the companion (Desidera et al. 2016) are worth further investigation.

\normalem
\begin{acknowledgements}
This research is based on the data obtained by the Subaru Telescope, 
operated by the National Astronomical Observatory of Japan.
This investigation has made use of the Kurucz database maintained by
Dr. R. L. Kurucz, and the VALD database operated at Uppsara University,
the Institute of Astronomy RAS in Moskow, and the University of Vienna. 
\end{acknowledgements}


\clearpage
\appendix

\section{Electronic data tables and figures}

Supplementary electronic materials (data tables and figure files) 
are accompanied with this article, which are separately contained 
in four directories as described below.   

\subsection{Atomic and molecular line data}

The directory ``linedat'' contains 34 files named as ``lines\_??????.dat'' 
(``??????'' is the 6-character region code; e.g., CH4273), which include
the data of atomic and molecular lines (typically several hundred lines) 
used for the fitting analysis at each region. 
The data are basically arranged in the ascending order of wavelength, 
though atomic and molecular lines are separately grouped in each file, 
Table~A.1 describes the contents of these line data.

\subsection{Data of observed and theoretical spectra}

In the directory ``specdat'' are contained 34 files named as ``fit\_??????.dat'' 
(``??????'' is the 6-character region code; e.g., CH4273), which include
the observed and fitted theoretical spectra at each region.
Each file consists of 119 sections corresponding to 118 program stars
and the Sun/Vesta (its number is tentatively designated as 999999). 
In each section, the first header line includes the information of
6-character HIP number (HIP), number of points ($n_{\rm p}$), first wavelength ($w_{1}$) 
and last wavelength ($w_{2}$), which can be read with the format (2X,A6,I4,2F10.4). 
Then, $\lambda$ (wavelength in \AA), $f_{\lambda}^{\rm obs}$
(observed spectra) $F_{\lambda}^{\rm the}$ (fitted theoretical spectra) are given 
with the format (F8.3,2F10.4) in each of the following $n_{\rm p}$ lines. 
Note that these spectra are the residual flux reduced to the theoretical 
continuum level ($F_{\rm cont}^{\rm the}$).  

\subsection{Figures of spectrum fitting}

The directory ``fitfigs'' contains 34 PDF files named as ``??????.pdf'' 
(``??????'' is the 6-character region code; e.g., CH4273), which include
the figures showing the accomplished fit between the observed (red open symbols)
and theoretical (blue lines) spectra for each of the 118(+1) stars,
which were constructed based on the ``fit\_??????.dat'' files.
Each spectrum (indicated by the corresponding HIP number) is vertically shifted 
by 0.5 relative to the adjacent ones. Note that these figures are arranged
in almost the same manner as adopted in our previous papers (e.g., fig.~8 in
Takeda et al. 2007 or fig.~4 in Takeda et al. 2011).

\subsection{Abundance results derived for each region}

Three data files ``relabs\_ch.dat'', ``relabs\_nh.dat'', and ``relabs\_oh.dat''
are found in the directory ``abunds'', which present the detailed results of 
relative abundances ([C/H] or [N/H] or [O/H] derived from each of the 11--12 
spectral regions)  and their means (along with the associated standard deviations 
and mean errors). Stellar parameters are also included for convenience. 
After the first header line, the results for each of the 118 stars are given 
in the 2nd through 119th lines. And the last 120th line is for the Sun/Vesta, 
where the absolute abundances [$A$(C) or $A$(N) or $A$(O)] resulting from 
each region (used as the reference abundances) are presented.  
The data contents and their format are described in Table~A.2.

\newpage

\begin{table*}[h]
\small
\caption{Contents of line data files (``lines\_??????.dat'').}
\begin{center}
\begin{tabular}
{ccccl}\hline\hline
 Bytes &  Format & Units &  Item &  Brief Explanations \\
\hline
~1--~9 &  F9.3 & \AA\  &   $\lambda$&  (air) wavelength \\
11--16 &  F6.2 &  ---  &    s-code    &  species code$^{(a)}$\\ 
18--20 &   A3  &  ---  &    species   &  notation of species$^{(b)}$\\ 
21--32 &  E12.4&  ---  &    $\eta$    & line-strength indicator$^{(c)}$\\
33--43 &  F11.5&  ---  &    $\eta$    & line-strength indicator$^{(c)}$\\ 
44--51 &  F8.3 &  eV   &   $\chi_{\rm low}$   &  lower excitation potential\\
52--59 &  F8.3 &  dex  &   $\log gf$  & log of stat. weight (lower level) times osc. strength\\
60--67 &  F8.3 &  dex  &   Gammar   &  radiation damping parameter$^{(d)}$ \\
68--75 &  F8.3 &  dex  &   Gammas   &  Stark effect damping parameter$^{(d)}$\\
76--83 &  F8.3 &  dex  &   Gammaw   &  van der Waals effect damping parameter$^{(d)}$\\
\hline
\end{tabular}
\end{center}
\small
Notes:\\
$^{(a)}$Constructed from the atomic number and ionization stage. For example:
     O~{\sc i} line  $\rightarrow$ 8.00, 
     Fe~{\sc i} line $\rightarrow$ 26.00, 
     Y~{\sc ii} line $\rightarrow$ 39.01,
     CH line $\rightarrow$ 106.00,
     NH line $\rightarrow$ 107.00,
     OH line $\rightarrow$ 108.00.\\
$^{(b)}$For example, Fe1 $\rightarrow$ Fe~{\sc i}, Y2 $\rightarrow$ Y~{\sc ii}.\\
$^{(c)}$Line-center-to-continuum opacity ratio calculated for the solar model atmosphere 
       (with the solar abundances) at $\tau_{5000} = 0.2$\\
$^{(d)}$Gammar: logarithm of radiation damping width (s$^{-1}$) [$\log\gamma_{\rm rad}$]. 
        Gammas: logarithm of Stark damping width (s$^{-1}$) per electron density (cm$^{-3}$) 
       at 10000 K [$\log(\gamma_{\rm e}/N_{\rm e})$].
       Gammaw: logarithm of van der Waals damping width (s$^{-1}$) per hydrogen density 
       (cm$^{-3}$) at 10000 K [$\log(\gamma_{\rm w}/N_{\rm H})$].
\end{table*}

\begin{table*}[h]
\small
\caption{Contents of abundance data files (``relabs\_?h.dat'').}
\begin{center}
\begin{tabular}
{ccccl}\hline\hline
 Bytes &  Format & Units &  Item &  Brief Explanations \\
\hline
~1--~6 &  I6  &  ---       & HIP    & Hipparcos catalogue number (999999 is for Sun/Vesta)\\
~7--13 & F7.0 &   K        & $T_{\rm eff}$   &  Effective temperature$^{(a)}$\\
14--19 & F6.2 &  dex       & $\log g$ &  Logarithm of surface gravity (in c.g.s.)$^{(a)}$\\
20--25 & F6.2 &  km~s$^{-1}$  &  $v_{\rm t}$ &     Microturbulent velocity dispersion$^{(a)}$\\
26--32 & F7.2 &  dex       & [Fe/H] &  Differential logarithmic Fe abundance relative to the Sun$^{(a)}$ \\
33--36 &  I4  &  ---       &  $n_{\rm t}$ & Total number of spectral regions \\
37--39 &  I3  &  ---       &  $n_{\rm s}$ & Number of regions adopted for calculation of mean $\langle$[X/H]$\rangle$\\
40--46 & F7.3 &  dex       &  $\langle$[X/H]$\rangle$ & Mean of [X/H]$^{(b)}$ averaged over different spectral regions\\  
47--52 & F6.3 &  dex       &  $\sigma$ & Standard deviation of $\langle$[X/H]$\rangle$\\
53--58 & F6.3 &  dex       &  $\epsilon$ & mean error of $\langle$[X/H]$\rangle$ ($\equiv \sigma/\sqrt{n_{\rm s}}$)\\
60--66 & F7.3 &  dex       &  [X/H]$_{1}$ & [X/H] value derived in region~1\\
67--67 &  A1  &  ---       &  flag$_{1}$ & Adopt-or-reject flag$^{(c)}$ for [X/H]$_{1}$\\  
$S_{i}$--$E_{i}$ &  F7.3 &  dex  &  [X/H]$_{i}$ & [X/H] value derived in region~$i$ $^{(d)}$\\
$F_{i}$--$F_{i}$ &   A1  &  ---  &  flag$_{i}$ & Adopt-or-reject flag$^{(c)}$ for [X/H]$_{i}$$^{(d)}$\\ 
\hline
\end{tabular}
\end{center}
\small
Notes:\\
$^{(a)}$These are the ``standard solutions'' derived in Takeda et al. (2007) (cf. sect.~3.1.1 therein).\\
$^{(b)}$[X/H] is the differential abundance of X (X is C or N or O) relative to the solar abundance; 
i.e., [X/H]~$\equiv$~$A_{*}({\rm X}) - A_{\odot}({\rm X})$\\
$^{(c)}$If the flag is 'x', this [X/H]$_{i}$ was judged to be anomalous (according to
Chauvenet's criterion) and excluded from the averaging process. Otherwise, this flag is blank.\\
$^{(d)}$$S_{i} = 52 + 8i$, $E_{i} = 58 + 8i$, and $F_{i} = 59 + 8i$, where $i$ is the region No. 
(ranging from 1 to $n_{\rm t}$).\\  
\end{table*}

\end{document}